%% file: main.tex
\documentclass[acmsmall,nonacm]{acmart}
\usepackage{amsfonts}
\usepackage{pifont}
\usepackage{textcomp}

\renewcommand\footnotetextcopyrightpermission[1]{}
\setcopyright{none}

\usepackage{amsthm}

\usepackage{makecell}

\usepackage{enumitem}
\usepackage{wrapfig}  % NOTE: not in TAPS list; used for wrapfigure environments

\theoremstyle{definition}

\newtheorem{thm3}{Definition}[section]
\newtheorem{definition}[thm3]{Definition}

\newtheorem{thm5}{Example}[section]
\newtheorem{example}[thm5]{Example}

\hypersetup{
	colorlinks=true,
	linkcolor=blue,
	citecolor=violet,
	filecolor=magenta, 
	urlcolor=blue,
}

\usepackage[linesnumbered,commentsnumbered,ruled,vlined]{algorithm2e}
\usepackage{stmaryrd}
\usepackage{listings}
\usepackage{semantic}
\usepackage{mathtools}
\usepackage{url}
\usepackage{subcaption}

\usepackage{bm}

\usepackage[utf8]{inputenc}
\usepackage{xspace}

\usepackage{cleveref}
\crefname{section}{§}{§§}
\Crefname{section}{§}{§§}

\usepackage{multirow}
\usepackage{fancybox}

\definecolor{codeblack}{rgb}{0,0,0}
\definecolor{codegreen}{rgb}{0,0.6,0}
\definecolor{codegray}{rgb}{0.5,0.5,0.5}
\definecolor{codepurple}{rgb}{0.58,0,0.82}
\definecolor{backcolour}{rgb}{1,1,1}

\definecolor{myblue}{RGB}{0, 0, 204}
\definecolor{mygreen}{RGB}{0, 129, 0}
\definecolor{mypurple}{RGB}{127, 0, 255}

\lstdefinestyle{mystyle}{
	backgroundcolor=\color{backcolour},
	commentstyle=\color{mygreen},
	language={Java}, 
	basicstyle=\ttfamily\scriptsize,
	keywordstyle=\scriptsize\bfseries\color{black},
	keywordstyle = [2]{\scriptsize\bfseries\color{myblue}},
	keywordstyle = [3]{\scriptsize\bfseries\color{mypurple}},
	keywordstyle = [4]{\scriptsize\bfseries\color{teal}},
	otherkeywords = {},
	morekeywords = [2]{ArrayList, LinkedHashMap, HashSet, LinkedList, TreeSet, HashMap, TreeMap, LinkedHashSet, List, Map, Set, Vector},
	morekeywords = [3]{put, add, get, contains, indexOf, remove, toString, size},
	breakatwhitespace=false,           
	captionpos=b,
	numbers=none,
	showspaces=false,                
	showstringspaces=false,
	showtabs=false, 
	tabsize=2
}

\newcommand{\blue}[1]{#1}
\newcommand{\bluetable}[1]{#1}
\newcommand{\del}[1]{}

\newcommand{\RNum}[1]{\uppercase\expandafter{\romannumeral #1\relax}}
\newcommand{\ToolName}{\textsc{NESA}}
\AtBeginDocument{%
  }

\begin{document}

%%
%% The "title" command has an optional parameter,
%% allowing the author to define a "short title" to be used in page headers.
\title{\ToolName: Relational Neuro-Symbolic Static Program Analysis}

%%
%% By default, the full list of authors will be used in the page
%% headers. Often, this list is too long, and will overlap
%% other information printed in the page headers. This command allows
%% the author to define a more concise list
%% of authors' names for this purpose.

%%
\author{Chengpeng Wang}
%\authornote{with author2 note}          %% \authornote is optional;
%% can be repeated if necessary
\orcid{0000-0003-0617-5322}
\affiliation{
	%\position{Position2a}
	\department{Department of Computer Science}             %% \department is recommended
	\institution{Purdue University}           %% \institution is required
	%\streetaddress{Street2a Address2a}
	\city{West Lafayette}
	%\state{State2a}
	%\postcode{Post-Code2a}
	\country{USA}                   %% \country is recommended
}
\email{wang6590@purdue.edu}         %% \email is recommended

\author{Yifei Gao}
%\authornote{with author2 note}          %% \authornote is optional;
%% can be repeated if necessary
\orcid{0000-0001-5421-4619}             %% \orcid is optional
\affiliation{
	%\position{Position2a}
	\department{Department of Computer Science}             %% \department is recommended
	\institution{Purdue University}           %% \institution is required
	%\streetaddress{Street2a Address2a}
	\city{West Lafayette}
	%\state{State2a}
	%\postcode{Post-Code2a}
	\country{USA}                   %% \country is recommended
}
\email{gao749@purdue.edu}         %% \email is recommended

\author{Wuqi Zhang}
%\authornote{with author2 note}          %% \authornote is optional;
%% can be repeated if necessary
\orcid{0000-0001-8039-0528}             %% \orcid is optional
\affiliation{
	%\position{Position2a}
	\department{Department of Computer Science}             %% \department is recommended
	\institution{Purdue University}           %% \institution is required
	%\streetaddress{Street2a Address2a}
	\city{West Lafayette}
	%\state{State2a}
	%\postcode{Post-Code2a}
	\country{USA}                   %% \country is recommended
}
\email{research@troublor.xyz}         %% \email is recommended

\author{Xuwei Liu}
%\authornote{with author2 note}          %% \authornote is optional;
%% can be repeated if necessary
\orcid{0009-0000-5319-1160}             %% \orcid is optional
\affiliation{
	%\position{Position2a}
	\department{Department of Computer Science}             %% \department is recommended
	\institution{Purdue University}           %% \institution is required
	%\streetaddress{Street2a Address2a}
	\city{West Lafayette}
	%\state{State2a}
	%\postcode{Post-Code2a}
	\country{USA}                   %% \country is recommended
}
\email{liu2598@purdue.edu}         %% \email is recommended

\author{Jinyao Guo}
%\authornote{with author2 note}          %% \authornote is optional;
%% can be repeated if necessary
\orcid{0009-0004-2352-8307}             %% \orcid is optional
\affiliation{
	%\position{Position2a}
	\department{Department of Computer Science}             %% \department is recommended
	\institution{Purdue University}           %% \institution is required
	%\streetaddress{Street2a Address2a}
	\city{West Lafayette}
	%\state{State2a}
	%\postcode{Post-Code2a}
	\country{USA}                   %% \country is recommended
}
\email{guo846@purdue.edu}         %% \email is recommended

\author{Mingwei Zheng}
%\authornote{with author2 note}          %% \authornote is optional;
%% can be repeated if necessary
\orcid{0009-0003-6032-6045}             %% \orcid is optional
\affiliation{
	%\position{Position2a}
	\department{Department of Computer Science}             %% \department is recommended
	\institution{Purdue University}           %% \institution is required
	%\streetaddress{Street2a Address2a}
	\city{West Lafayette}
	%\state{State2a}
	%\postcode{Post-Code2a}
	\country{USA}                   %% \country is recommended
}
\email{zheng618@purdue.edu}         %% \email is recommended

\author{Qingkai Shi}
%\authornote{with author2 note}          %% \authornote is optional;
%% can be repeated if necessary
\orcid{0000-0002-8297-8998}             %% \orcid is optional
\affiliation{
	%\position{Position2a}
	\department{Department of Computer Science}             %% \department is recommended
	\institution{Purdue University}           %% \institution is required
	%\streetaddress{Street2a Address2a}
	\city{West Lafayette}
	%\state{State2a}
	%\postcode{Post-Code2a}
	\country{USA}                   %% \country is recommended
}
\email{shi553@purdue.edu}         %% \email is recommended

\author{Xiangyu Zhang}
%\authornote{with author2 note}          %% \authornote is optional;
%% can be repeated if necessary
\orcid{0000-0002-9544-2500}             %% \orcid is optional
\affiliation{
	%\position{Position2a}
	\department{Department of Computer Science}             %% \department is recommended
	\institution{Purdue University}           %% \institution is required
	%\streetaddress{Street2a Address2a}
	\city{West Lafayette}
	%\state{State2a}
	%\postcode{Post-Code2a}
	\country{USA}                   %% \country is recommended
}
\email{xyzhang@purdue.edu}         %% \email is recommended

\renewcommand{\shortauthors}{Chengpeng Wang et al.}

\input{abstract}




\maketitle

\input{introduction.tex}
\input{motivation.tex}

\input{reduction.tex}

\input{approach.tex}
\input{evaluation.tex}
\input{related_work.tex}
\input{conclusion.tex}

\bibliographystyle{ACM-Reference-Format}
\bibliography{sample-base}

%%
%% If your work has an appendix, this is the place to put it.
\appendix
\end{document}

%% file: abstract.tex
\begin{abstract}
Static program analysis plays an essential role in program optimization, bug detection, and debugging. However, reliance on compilation and limited customization hinder its adoption in the real world. This paper presents a compositional neuro-symbolic approach named NESA that facilitates compilation-free and customizable static program analysis using large language models (LLMs) with mitigated hallucinations. Specifically, we propose an analysis policy language, a restricted form of Datalog, to support users decomposing a static program analysis problem into several sub-problems that target simpler syntactic or semantic properties upon smaller code snippets. The problem decomposition enables the LLMs to target more manageable semantic-related sub-problems with reduced hallucinations, while the syntactic ones are resolved by parsing-based analysis without hallucinations. An analysis policy then is evaluated with lazy and incremental prompting, which significantly mitigates the hallucinations and improves the performance. We evaluate NESA for program slicing and bug detection upon benchmark and real-world programs. Evaluation results show that while NESA supports compilation-free and customizable analysis, it can still achieve comparable and even better performance than existing techniques. In a customized taint vulnerability detection upon TaintBench, for example, NESA achieves a precision of 66.27\%, a recall of 78.57\%, and an F1 score of 0.72, surpassing an industrial approach by 0.20 in F1 score. NESA also detects 13 real-world memory leak bugs, which have been fixed by developers.
\end{abstract}

%% file: introduction.tex
\section{Introduction}
\label{sec:intro}

Static program analysis has long been an essential technique in the software development, facilitating various software engineering tasks such as program optimization~\cite{DBLP:journals/infsof/Reps98, DBLP:conf/pldi/IkarashiBRGR22}, bug detection~\cite{CalcagnoDOY09, Arzt14FlowDroid}, and repair~\cite{DBLP:conf/icse/NguyenQRC13, DBLP:journals/pacmse/Song0LCR24}. Despite significant progress in precision, efficiency, and scalability over recent decades~\cite{Xue16SVF, Shi18Pinpoint}, its widespread adoption in the industry has lagged behind expectations.
According to existing studies~\cite{DBLP:conf/icse/JohnsonSMB13, ChristakisB16, DBLP:conf/asplos/ZhangCYWTZ24}, two significant limitations contribute to the underutilization of static program analysis.
First, mainstream static program analyzers~\cite{Arzt14FlowDroid, CalcagnoDOY09} often rely on the compilation process from source code to intermediate representations (IRs), limiting their usability in development environments where code is incomplete. 
Second, existing analyzers only provide limited support for customization for specific requirements of developers~\cite{ChristakisB16},
which requires experienced compiler hacking skills and expertise of IRs.
These two limitations erect significant barriers, discouraging practitioners from fully integrating analyzers into development workflows.

In the past few years, LLMs have shown their exceptional performance in programming-related tasks~\cite{DBLP:conf/emnlp/ZhangCZKLZMLC23, LLMRepair, JimenezYWYPPN24, LingmingFuzz}.
The capability of understanding program semantics makes LLMs stand out as a promising alternative to static program analyzers~\cite{DBLP:journals/corr/abs-2401-00812}. 
With a bug definition and buggy examples, 
LLMs can identify bugs in code snippets without compilation. 
Unlike traditional static program analyzers,
this prompting-based approach demonstrates another attractive paradigm that naturally supports compilation-free and customizable analysis.
However, inherent hallucinations~\cite{Shi23} make LLMs the \emph{Sword of Damocles} when solving specific static program analysis problems,
potentially resulting in false positives or negatives.

To mitigate hallucinations, our approach is grounded in two key insights.
First, a sophisticated analysis problem can be decomposed into smaller, more tractable sub-problems focusing on simpler program properties within compact code snippets, which narrows the program scope and simplifies the properties that LLMs address, thereby reducing the risk of hallucinations.
As such, we introduce a Datalog rule, e.g., $R_1(x, y) \leftarrow R_2(x, z), R_3(z, y)$, which enables the users to decompose the original property (e.g., the one depicted by $R_1(x, y)$) into several simpler properties (e.g., the ones depicted by $R_2(x, z)$ and $R_3(z, y)$).
Second, while semantic analysis is generally undecidable~\cite{rice1953classes}, many syntactic properties can be effectively addressed through deterministic, parsing-based analysis. 
Hence, we can analyze semantic properties via prompting while leveraging a parser to derive syntactic ones, thereby removing hallucinations in the syntactic analysis.

Based on the insights, we propose the relational neuro-symbolic static program analysis \ToolName.
\del{Specifically, we introduce two kinds of relations in the Datalog-like analysis policy to represent syntactic and semantic properties, respectively. Also, we develop an evaluation procedure that evaluates the analysis policy by applying a program parser and user-customized LLM prompting.}
\blue{The novelty of \ToolName\ is twofold.
First, we introduce a Datalog-like policy language incorporating two kinds of relations, namely symbolic and neural relations, which enables the principled composition of syntactic and semantic analysis results. 
Second, we propose an evaluation procedure that evaluates the analysis policy by applying a program parser and user-customized LLM prompting.}
Technically, the evaluation procedure benefits from two innovative designs. 
First, we introduce the \emph{lazy prompting} in evaluating each Datalog rule,
effectively reducing the hallucinations in populating neural relations.
Second, we propose the \emph{incremental prompting} to skip unnecessary prompting rounds,
ensuring the tuples in the neural relations would not be generated by LLMs multiple times.

\begin{figure*}
\centering
\includegraphics[width=0.96\textwidth]{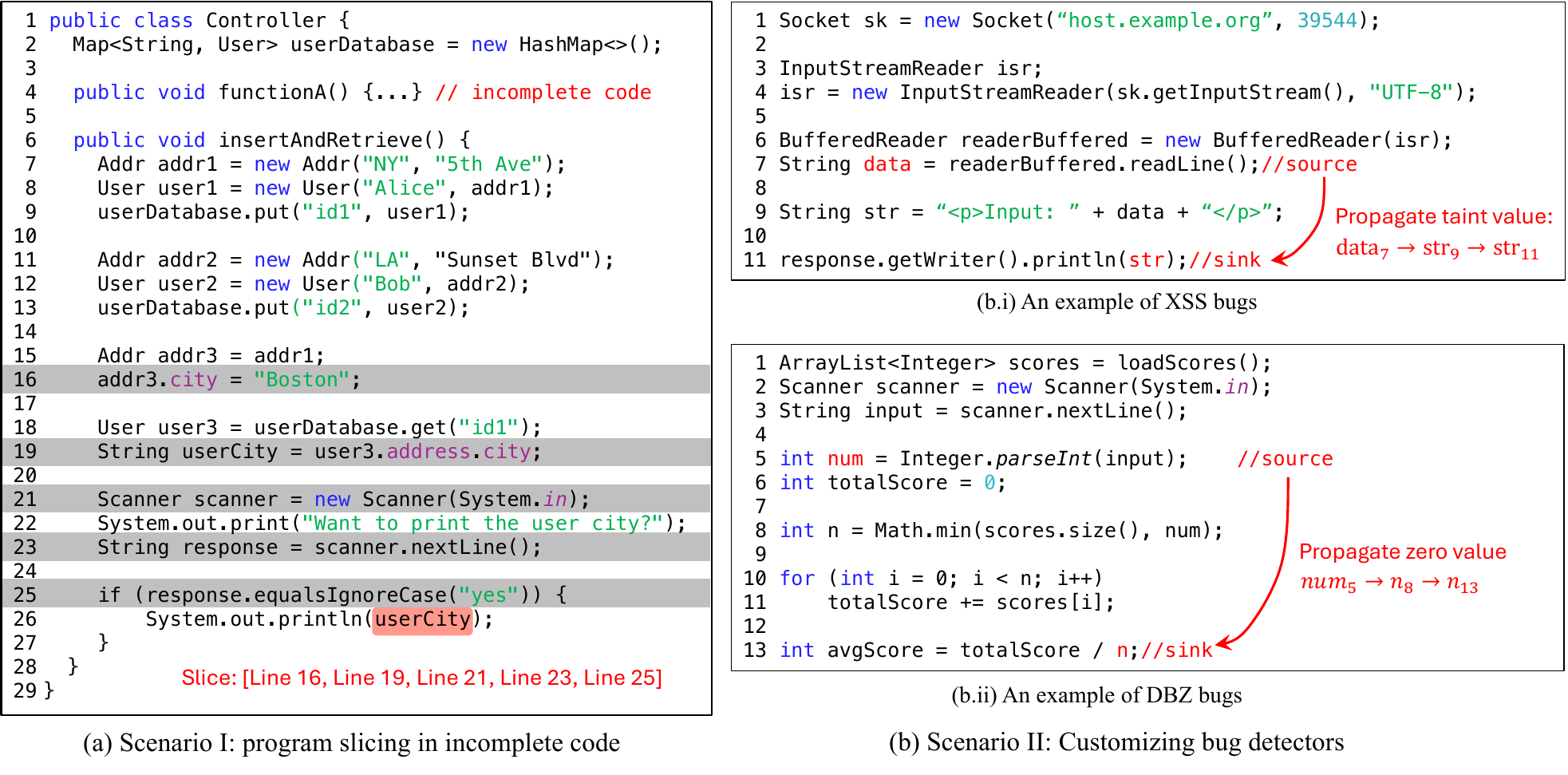}
% \vspace{-2mm}
\caption{Two motivating examples of compilation-free and customizable static program analysis}
\label{fig:example}
\vspace{-9mm}
\end{figure*}

We have developed a prototype of \ToolName~\cite{nesa_artifact} and extensively evaluated it across program slicing and bug detection. 
It is shown that \ToolName\ achieves 91.50\% precision and 84.61\% recall in the program slicing,
surpassing the state-of-the-art program slicer \textsc{NS-Slicer}~\cite{DBLP:journals/pacmpl/YadavallyLWN24} by 0.06 F1 score.
\blue{The average precision and recall of \ToolName\ in the bug detection upon Juliet Test Suite~\cite{BolandB12} reach 86.84\% and 87.69\%, respectively.}
In real-world malware applications within TaintBench~\cite{LuoPPBPMBHM22}, \ToolName\ successfully detects 55 out of 70 taint vulnerabilities, achieving a precision of 66.27\% and a recall of 78.57\%, which surpasses an industrial tool by 37.14\% recall and 0.20 F1 score.
\ToolName\ also uncovers 13 memory leak bugs in real-world programs, all of which have been confirmed and fixed by the developers.
\blue{Unlike existing techniques, \ToolName\ offers a general analysis framework, which requires little manual labor and expert knowledge in the customization. On average, an analysis policy contains 11.1 Datalog rules, 5.6 symbolic relations, and 2.7 neural relations, and meanwhile, a prompt file only spans approximately 43.4 lines in a JSON file.}
To summarize, the key contributions include:
\begin{itemize}[leftmargin=*]
\item We propose a relational neuro-symbolic approach that utilizes LLMs to enable compilation-free and customizable analysis.

\item We present a series of evaluation strategies, including lazy and incremental prompting, to mitigate the innate hallucinations of LLMs and significantly improve the performance.

\item We perform an extensive group of experiments to show that \ToolName\ achieves comparable and even superior performance relative to the specialized SOTA targeting specific tasks.
\end{itemize}

%% file: motivation.tex
\section{Overview}
\label{sec:overview}

In this section, we first summarize the dilemma of current static analysis techniques (\cref{subsec:dilemma}), motivate the prompting-based static analysis (\cref{subsubsec:sap}), and outline our overarching idea (\cref{subsubsec:approach}).

\subsection{The Dilemma of Static Program Analysis}
\label{subsec:dilemma}
Static analysis, a technique that analyzes code without execution, has been extensively researched for several decades but has not been as widespread as expected. In what follows, we highlight two
critical factors that cause the dilemma and motivate the new static analysis paradigm in this work.

\subsubsection{Reliance of Compilation}
\label{subsubsec:compilation}

Mainstream analyzers, such as \textsc{FlowDroid}~\cite{Arzt14FlowDroid} and \textsc{Infer}~\cite{CalcagnoDOY09}, are composed of sophisticated semantic analyses applied to IR code generated during the compilation. However, these tools fail to analyze incomplete programs, especially the ones in development environments. As shown in Fig.~\ref{fig:example}(a), the class \textsf{Controller} is incomplete during development. The developer may need a program slicer~\cite{DBLP:conf/pldi/SridharanFB07} to localize the statements affecting the variable \textsf{userCity} at line 26. The program slicer should identify alias pairs, such as $(\textsf{addr1}, \textsf{addr3})$ and $(\textsf{user1}, \textsf{user3})$, eventually extracting a program slice that contains lines 16, 19, 21, 23, and 25. However, the absence of IR code on incomplete programs hinders the effectiveness of these analyzers.

Even when a program is compilable, most static program analyzers have to interfere with the compilation configuration to access IR code~\cite{DBLP:conf/asplos/ZhouYHCZ24}. For example, many LLVM-IR-based analyzers~\cite{Xue16SVF, Shi18Pinpoint} require C/C++ projects to be compiled using specific compiler wrappers~\cite{WLLVM}, while C/C++ languages support over 20 build systems and 36 compilers, making it extremely difficult and error-prone to accommodate such a vast array of native build systems and compilers. What's even worse, the ongoing evolution of IR versions compounds the problem. Many projects may require newer compilers to generate higher-version IR code, while many static program analyzers are compatible only with specific lower IR versions. This mismatch creates an IR version trap~\cite{DBLP:conf/asplos/ZhangCYWTZ24}, impeding the widespread adoption of static program analyzers.

\subsubsection{Lack of Customization Support}
\label{subsubsec:customization}

According to a survey by Microsoft~\cite{ChristakisB16}, 21\% of developers ceased using static program analyzers because the analyzers could not meet specific needs. For instance, \textsc{FlowDroid}~\cite{Arzt14FlowDroid} detects taint-style bugs, such as the Cross-Site Scripting (XSS) bug in Fig.~\ref{fig:example}(b.i), by tracing taint flows from sources (i.e., the origins of malicious values) to sinks (i.e., the operands of dangerous operations), where the values of sinks depend on the value of sources. For other bugs like the Divide-by-Zero (DBZ) bug in Fig.~\ref{fig:example}(b.ii), \textsc{FlowDroid} fails to support detection where the sink's value is identical to the source, rather than merely dependent on it.

To perform specialized analyses, developers must either build new static program analysis tools from scratch or modify existing ones.
It is reported that over 70\% of developers who previously used static program analyzers are dissatisfied with existing tools because they do not accommodate desired customization~\cite{DBLP:conf/icse/JohnsonSMB13}. Specifically, customizing static program analyzers also requires in-depth knowledge of compiler infrastructures, thereby significantly raising the barrier to tool usage. Although recent tools like \textsc{CodeQL}~\cite{AvgustinovMJS16} enable customization through queries, developers are still required to learn a domain-specific language with an extensive set of APIs, often introducing a steep learning curve.

\subsection{Compilation-free and Customizable Static Program Analysis}
\label{subsec:promptingsa}

To fill the research gap, we propose a compilation-free and customizable static analysis. In what
follows, we will first introduce static analysis via prompting, then highlight the hallucination issues,
and lastly outline the key idea of our solution.

\subsubsection{Static Program Analysis via Prompting}
\label{subsubsec:sap}

LLMs have demonstrated remarkable performance in understanding program semantics~\cite{KexinICML23, DBLP:journals/corr/abs-2312-08477, DBLP:journals/corr/abs-2308-03312, DBLP:conf/pldi/BrentGLSS20},
suggesting that they can be utilized to analyze code without compilation.
Users can provide the definitions of program slices and bug types along with examples in the few-shot chain-of-thought (CoT) prompting~\cite{CoT}. Then the LLMs would output a program slice or bugs. Unlike modifying existing analyzers, users can specify the analysis task in natural languages and offer examples, which requires no expertise in compiler internals.

As shown by existing efforts in the NLP community~\cite{JiYXLIF23, Shi23}, LLMs have become the \emph{Sword of Damocles} in solutions to many domain problems due to their inherent hallucinations. The similar challenge also exists in prompting-based static program analysis.
For instance, the precision and the recall of the DBZ detection powered by GPT-3.5-Turbo were both lower than 5\% when we applied few-shot CoT prompting to Java programs in Juliet Test Suite~\cite{BolandB12}, as the model fails to precisely identify potential zero values and track their propagation. 

\subsubsection{Our Approach}
\label{subsubsec:approach}
In this paper, we propose a systematic solution to mitigate the hallucinations in prompting-based static analysis,
which eventually facilitates a compilation-free and customizable analysis with exceptional performance.
Our key ideas originate from two key observations:
\begin{itemize}[leftmargin=*]
	 \item A static analysis problem can be decomposed into sub-problems that focus on simpler program properties within smaller code snippets. For example, program slicing can be reduced to analyzing data and control dependencies of specific program values. Similarly, the XSS and DBZ bug detection can be divided into source/sink extraction and specific forms of taint flow reachability analysis. By addressing these more manageable sub-problems, LLMs are less likely to introduce hallucinations during prompting.
	
	\item Although non-trivial semantic analysis problems are generally undecidable~\cite{rice1953classes}, there still exists a wide range of syntactic program properties that can be deterministically solved by a parsing-based analysis. In program slicing, for example, we can easily collect all the conditions that guard a specific program line by parsing and further determine the control dependencies. By decoupling syntactic properties from semantic ones, we can effectively avoid LLM hallucinations when solving syntactic-related sub-problems.
\end{itemize}

\begin{figure}[t]
	\centering
	\includegraphics[width=0.8\textwidth]{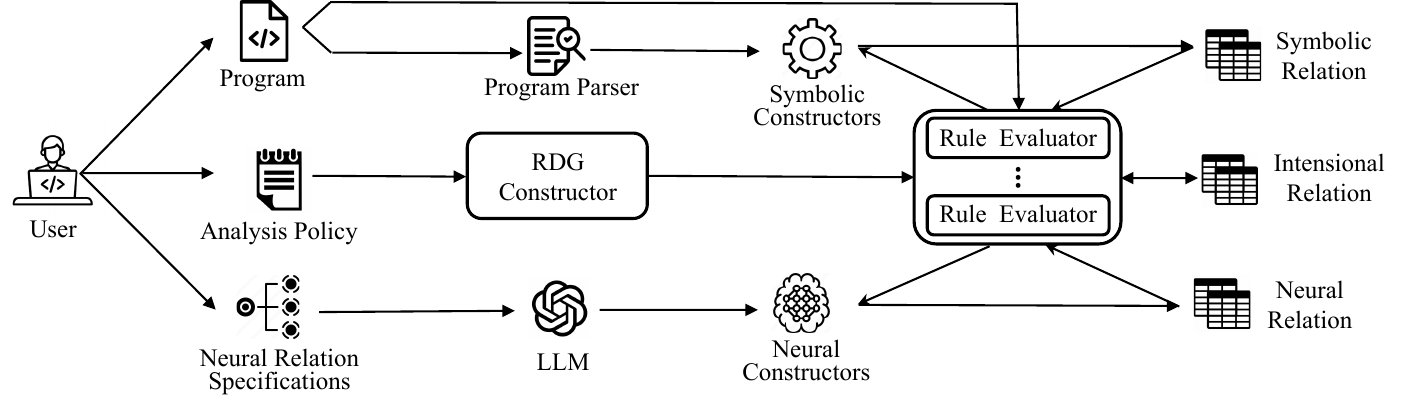}
	\caption{The workflow of \ToolName}
    \vspace{-5mm}
	\label{fig:workflow}
\end{figure}

\begin{figure}[t]
	\centering
	\includegraphics[width=\textwidth]{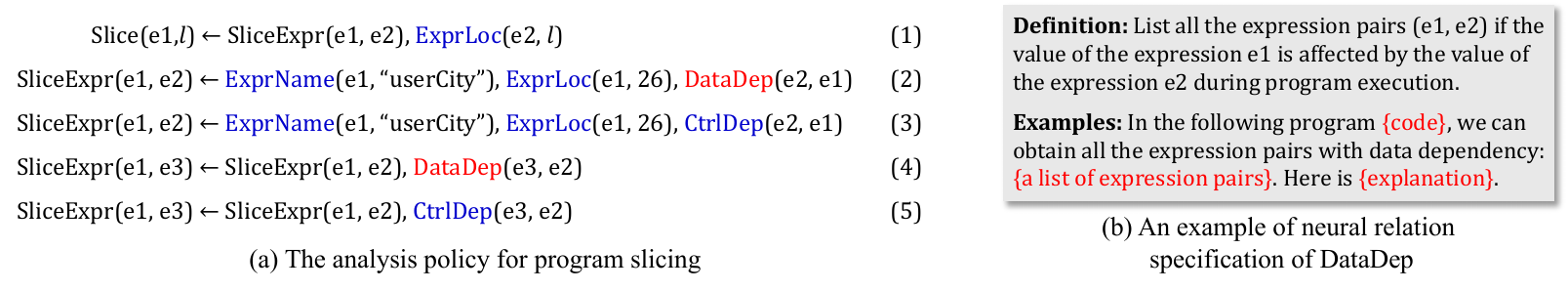}
	\caption{The examples of analysis policy and neural relation specification. In the sub-figure (a), the symbolic, neural, and intensional relations are in {\color{myblue}{blue}}, {\color{red}{red}}, and black, respectively.}
    \vspace{-5mm}
	\label{fig:input}
\end{figure}

Based on these insights, we present \ToolName, a compositional neuro-symbolic approach that supports compilation-free and customizable analysis, of which the workflow is demonstrated in Figure~\ref{fig:workflow}.
Apart from the analyzed program, the inputs of \ToolName\ also include:
\begin{itemize}[leftmargin=*]
	\item \emph{Analysis policy} that specifies the problem decomposition. Specifically, we employ a restricted form of Datalog as our analysis policy language. It allows users to introduce \emph{symbolic} and \emph{neural relations}, which depict syntactic and semantic properties as prerequisites, and derive desired program properties depicted by \emph{intensional relations} in Datalog rules.
	For example, Figure~\ref{fig:input}(a) shows the analysis policy of program slicing with the slicing seed \textsf{userCity} at line 26. 
	The symbolic relations \textsf{ExprName} and \textsf{ExprLoc} support localizing the slicing seed, while
	the symbolic relation \textsf{CtrlDep} and the neural relation \textsf{DataDep} serve as the ingredients for slicing.
	
	\item \emph{Neural relation specifications} that determine how to derive semantic properties depicted by neural relations via prompting. 
	They consist of the definitions of semantic properties and several examples with explanations.
	In program slicing, the specified neural relation specification in Figure~\ref{fig:input}(b) determines how to populate the neural relation \textsf{DataDep} depicting data dependencies.
\end{itemize}

Using these inputs, \ToolName\ achieves customized static analysis by evaluating the specified analysis policy. Based on the worklist algorithm, \ToolName\ can obtain the output relation at the fixed point, which depicts the desired program property. Specifically, we first construct a \emph{rule dependency graph} (RDG) %\shi{example?} 
for the analysis policy, which encodes the dependency relation between rules and guides the iterative fixed point computation.
To evaluate a Datalog rule in the analysis policy, we need to determine the contents of the symbolic and neural relations,
which are prerequisites for deriving the program property depicted by the intensional relation in the Datalog rule.
Technically,
\ToolName\ uses a program parser to generate tuples in symbolic relations
and selects appropriate \textit{neural constructors}, which are instantiated by user-specified neural relation specifications, to populate neural relations.
However, populating neural relations with LLMs can still induce hallucinations, yielding low-quality analysis results.
Also, each prompting round may have non-trivial time and token costs, which can become unacceptable when the number of prompting rounds is large.
To address these challenges, we introduce two essential technical designs:

\begin{itemize}[leftmargin=*]
\item \emph{Lazy prompting}: During rule evaluation, we delay populating the neural relation with prompting until the contents of other non-neural relations are also determined. In the evaluation of the rule (2) in Figure~\ref{fig:input}(a), for example, we first determine the contents of the symbolic relations \textsf{ExprName} and \textsf{ExprLoc} and then join their tuples to narrow down the possible values of $\textsf{e1}$. In this way, we can enforce LLMs to focus on restricted program expressions in the prompting when, which would mitigate the hallucinations of the population of the neural relation \textsf{DataDep}.

\item \emph{Incremental prompting}: As rules may be evaluated multiple times before reaching a fixed point, tuples for neural relations must be generated iteratively. To avoid prompting redundantly, we apply LLM-based neural relation constructors to generate each tuple in the neural relation only once, effectively reducing token costs. For example, we only generate the pairs of the expressions with data dependencies one time and store them in the relation \textsf{DataDep} in Figure~\ref{fig:input}(a).
\end{itemize}

In what follows, Section~\ref{sec:language} defines the analysis policy language. Section~\ref{sec:approach} discusses the evaluation procedure achieving compilation-free and customizable analysis.

%% file: reduction.tex
\section{Analysis Policy Language}
\label{sec:language}

This section introduces the analysis policy language and two important relations.
Lastly, we formulate a static program analysis problem as the analysis policy evaluation.

\subsection{Syntax}
\label{subsec:syntax}

\begin{wrapfigure}[13]{r}{0.48\textwidth}
\vspace{-12mm}
\centering
\scalebox{1.0}{
	\begin{minipage}{0.48\textwidth}
		\centering
		\begin{figure}[H]
			\footnotesize
			\centering
			\[
			\begin{aligned}
			\text{Analysis Policy} \quad  &p ::= r_1 \mid r_2 \mid \dots \mid r_n \\
			\text{Rule} \quad             &r ::= R_I \leftarrow R_1, R_2, \dots, R_n \\
			\text{Relation} \quad         &R ::= R_S \mid R_N \mid R_I \\
			\text{Symbolic Relation} \quad    &R_S ::= R_S(a_1) \mid R_S(a_1, a_2) \\
			\text{Neural Relation} \quad      &R_N ::= R_N(a_1) \mid R_N(a_1, a_2) \\
			\text{Intensional Relation} \quad &R_I ::= R_I(a_1) \mid R_I(a_1, a_2) \\
			\text{Term} \quad & a \in \textit{Dom} ::= \textit{Expr} \cup \textit{String} \cup \textit{Int}
			\end{aligned}
			\]
			\vspace{-2mm}
			\caption{The syntax of analysis policy}
			\vspace{-9mm}
			\label{fig:syntax}
		\end{figure}
		\begin{figure}[H]
			\footnotesize
			\centering
			\begin{align}
			\text{XSS}(\text{e1}, \text{e2}) & \leftarrow \text{Src}(\text{e1}), \text{\color{red}{TaintProp}}(\text{e1}, \text{e2}), \text{Sink}(\text{e2})\\
			\text{Src}(\text{e1}) &\leftarrow \text{\color{red}{XSSSrcNeural}}(\text{e1})\\
			\text{Sink}(\text{e1}) &\leftarrow \text{\color{red}{XSSSinkNeural}}(\text{e1})
			\end{align}
			\vspace{-4mm}
			\caption{An example of analysis policy}
			\label{fig:xss-rules}
		\end{figure}
	\end{minipage}
}
\end{wrapfigure}
As shown in Fig.~\ref{fig:syntax},
\emph{an analysis policy is a set of Datalog rules that solve a specific static program analysis problem via divide-by-conquer.}
Each Datalog rule derives a specific relation in its \emph{head} (i.e., its left hand), namely an \emph{intensional relation}, from the relations in its \emph{body} (i.e., its right hand).
Unlike traditional Datalog,
we introduce \emph{symbolic} and \emph{neural} relations depicting syntactic and semantic properties, respectively,
which appear in the body of a rule.
An \emph{output relation} is an intensional relation that indicate the desired property and can not appear in the body of the rules which do not derive it.
In this work, we only consider unary and binary relations and target expression-related properties, which can formulate most common program properties in static program analysis.
Hence, the domain of the terms is set to contain expressions, string values, and integer.
Here, string values can encode the string representations of expressions,
while integer values indicate program lines.
Without the loss of expressiveness, we restrict the number of neural relations in the rule to be at most one.
Other rules can be transformed into multiple rules conforming to our syntax by introducing intermediate intensional relations.

\begin{example}
Fig.~\ref{fig:input}(a) shows the analysis policy of program slicing. 
The rules (2)--(5) collect data and control dependencies. The output relation \textsf{Slice} derived by the rule (1) indicates the desired slice.
In Fig.~\ref{fig:xss-rules}, intra-procedural XSS bug detection is decomposed into identifying sources/sinks with rules (2) and (3) and determining intra-procedural taint flows with rule (1). 
\end{example}

\subsection{Symbolic and Neural Relations}
\label{subsec:sr}
Symbolic and neural relations are the basis of the analysis,
depicting syntactic and semantic properties as prerequisites.
In what follows, we will provide more details on them.

\subsubsection{Symbolic Relation}
In static program analysis, we basically focus on the properties of specific constructs. For instance, specific expressions, including those associated with specific names or located at specific lines, function parameters/return values, and arguments/output values at function call sites, are often the pivot constructs that facilitate the analysis. Besides, program properties relating to control flows, such as control-flow order and control dependencies, are also required in many static program analysis applications. These program properties can be directly obtained from the abstract syntax tree, offering broad applicability across various static program analysis tasks.

To depict common syntactic properties, we introduce a set of symbolic relations shown in Table~\ref{tab:symbolic-relation-syntax}.
Specifically, the relations \textsf{Args} and \textsf{Outs} maintain the arguments and outputs of function calls, respectively,
while the relations \textsf{Paras} and \textsf{Rets} maintain the parameters and return values of functions, respectively.
To localize specific expressions, we offer two relations, namely \textsf{ExprName} and \textsf{ExprLoc}, that maintain the expressions along with their names and program lines, respectively.
We also introduce the relations \textsf{CtrlOrd} and 
\textsf{CtrlDep} to maintain the control-flow order and 
control dependency between different expressions.
To support inter-procedural analysis, we introduce the relations \textsf{ArgPara} and \textsf{OutRet}
that match the arguments and outputs of function calls with the parameters and return values of callee functions, respectively.
Our work does not consider calling contexts and 
focuses on context-insensitive analysis, while it can be easily extended to context sensitive analysis by using additional relations depicting calling contexts.

\begin{table}[t]
\footnotesize
	\centering
	\caption{Symbolic relations and their descriptions}
	% \footnotesize
	\begin{tabular}{p{0.34\linewidth} p{0.58\linewidth}}
		\toprule
		\textbf{Symbolic Relation} & \textbf{Description} \\ \midrule
		\text{Args(e:Expr)} & Argument $\text{e}$ of a function call \\
		\text{Outs(e:Expr)} & Output value $\text{e}$ of a function call \\
		\text{Paras(e:Expr)} & Parameter $\text{e}$ of a function \\
		\text{Rets(e:Expr)} & Return value $\text{e}$ of a function \\
		\text{Exprs(e:Expr, s:Str)} & Expression $\text{e}$ with name $s$ \\
		\text{ExprLoc(e:Expr, $l$:Int)} & Expression $\text{e}$ at line $l$ \\ \midrule
		\text{CtrlOrd(\text{e1}:Expr, \text{e2}:Expr)} & Expression $\text{e1}$ may be evaluated before $\text{e2}$\\
		\text{CtrlDep(\text{e1}:Expr, \text{e2}:Expr)} & Expression $\text{e1}$ affects whether $\text{e2}$ is evaluated\\
		\text{ArgPara(\text{e1}:Expr, \text{e2}:Expr)} & Argument $\text{e1}$ match with parameter $\text{e2}$ \\
		\text{OutRet(\text{e1}:Expr, \text{e2}:Expr)} & Output value $\text{e1}$ matches with return value $\text{e2}$ \\ \bottomrule
	\end{tabular}
	\label{tab:symbolic-relation-syntax}
    \vspace{-3mm}
\end{table}

\begin{example}
As shown in Fig.~\ref{fig:input}(a), the symbolic relations  \textsf{ExprName} and \textsf{ExprLoc} 
enable the identification of the slicing seed $\textsf{userCity}$ at line 26 of Fig.~\ref{fig:example}(a), and \textsf{CtrlDep} collects the expressions which are the control dependencies of $\textsf{userCity}$.
\end{example}

\subsubsection{Neural Relation}

While symbolic relations capture syntactic properties, many semantic properties, such as data dependencies, cannot be easily extracted via parsing. To address this, our analysis policy incorporates neural relations for customized semantic analysis. 
% The key insight is that prompting LLMs offers a user-friendly interface for deriving semantic properties. 
By providing a property definition and several examples, the desired semantic property can be obtained via few-shot CoT prompting, thereby populating the neural relations. We formalize \emph{neural relation specification} as follows to formulate how a user defines a neural relation.

\begin{definition}(Neural Relation Specification)
	Given a neural relation, its neural relation specification is a pair of a natural language description $D$ and a set of examples $\mathcal{E}$. Here, the natural language description $D$ defines the desired semantic property. An example in $\mathcal{E}$ contains an example program $P_e$ and an explanation $E_e$ on the program property.
\end{definition}

\begin{example}
	\label{ex:nrs}
    Fig.~\ref{fig:input}(c) shows the neural relation specification of the neural relation \textsf{DataDep} used in Fig.~\ref{fig:input}(a). It can enforce the LLMs produce the tuples in \textsf{DataDep} depicting all the expressions with data dependencies.
\end{example}

As shown in Definition~\ref{ex:nrs}, the supplied definition and examples can coach LLMs in deriving semantic properties without compilation. More importantly, writing such specifications is declarative and inductive, eliminating the need for users to modify compiler internals for customization.

\subsection{Static Program Analysis as Analysis Policy Evaluation}
\label{subsec:pf}
Based on the symbolic relations in Table~\ref{tab:symbolic-relation-syntax} and customized neural relations with their specifications,
we can derive the targeted program property by evaluating the analysis policy.
Thus, we reduce a static program analysis problem to an analysis policy evaluation problem, formulated as follows.

\smallskip
\emph{\textbf{Problem Formulation.}}
Given a program $P$, an analysis policy $p$, and a set of neural relation specifications \emph{$\textit{Specs}$},
derive all the tuples of the output relation $R^{*}$ in the analysis policy $p$,
which depicts the targeted program property of the program $P$.

\smallskip
Unlike traditional Datalog, evaluating an analysis policy exhibits two unique challenges.
First, populating neural relations via prompting may introduce hallucinations, which cause the missing or wrong tuples in the neural relations,
further yielding the low precision and recall.
Second, the evaluation may require numerous prompting rounds, 
potentially consuming substantial token and time costs.
Section~\ref{sec:approach} will present the details of our relational neuro-symbolic static program analysis, which addresses the above challenges.

%% file: approach.tex
\section{Analysis Policy Evaluation}
\label{sec:approach}

This section presents our relational neuro-symbolic static program analysis via analysis policy evaluation.
We first provide an overview and then introduce two kinds of constructors,
followed by the illustration of the rule evaluation.

\subsection{Overview of the Evaluation Procedure}
\label{subsection:overview}
To solve the targeted static analysis problem,
we need to evaluate the analysis policy specified by users
with an instantiation of the worklist algorithm.
To facilitate the formalization, we first formally define two important concepts, namely \emph{analysis state} and \emph{rule semantics}, as follows.

\begin{definition}(Analysis State)
Given an analysis policy $p$, an analysis state $\mathbf{S}$ is a function that maps a relation symbol in $p$ to a set of tuples belonging to the relation.
\end{definition}

\begin{definition}(Rule Semantics)
\label{def:rs}
Given a set of neural relation specifications \emph{\textsf{Specs}} and a rule $r$ in an analysis policy $p$, the rule semantics $\llbracket r \rrbracket_{\emph{\textsf{Specs}}}$ is a function that maps a pair of a program and an analysis state $\mathbf{S}$ to another analysis state $\mathbf{S}'$, where $\mathbf{S}$ and $\mathbf{S}'$ intuitively indicate the content of the relations before and after applying the rule $r$, respectively.
\end{definition}

\begin{example}
\label{ex:state_semantics}
Consider the rule (3) in Figure~\ref{fig:input}(a) and the program $P$ in Figure~\ref{fig:example}(a). Initially, we assume $\mathbf{S}(R) = \emptyset$ for any relation symbol $R$.
By applying rule (3), we obtain a new analysis state $\mathbf{S}' = \llbracket r_3 \rrbracket_{\emph{\textsf{Specs}}}(P, \mathbf{S})$, 
where $\mathbf{S'}(\textsf{ExprName}) = \{ (\textsf{userCity}_{19}, \  \textsf{``userCity''}), \  (\textsf{userCity}_{26}, \ \textsf{``userCity''})\}$,  $\mathbf{S'}(\textsf{ExprLoc}) = \{ (\textsf{userCity}_{26}, 26)\}$,
$\mathbf{S'}(\textsf{CtrlDep}) = \{(\textsf{response}_{25}, \textsf{userCity}_{26}) \}$, and $\mathbf{S'}(\textsf{SliceExpr}) = \{  (\textsf{userCity}_{26}, \ \textsf{response}_{25}) \}$.
Here, $\textsf{userCity}_{19}$ refers to the expression \textsf{userCity} at line 19, and the same applies to other similar notations.
\end{example}

The analysis state intuitively depicts the discovered tuples indicating specific program properties,
while the rule semantics determines how tuples are derived from existing ones.
Notably, the tuples of specific relations may further be utilized to derive the tuples of other relations.
To formalize this dependency relation,
we introduce the \emph{rule dependency graph (RDG)} as follows to show the dependencies between the rules,
which can further guide our evaluation procedure.

\begin{definition}(Rule Dependency Graph)
	Given an analysis policy $p$, its rule dependency graph $G_r$ is a pair $(V, E)$ where $V$ contains all the rules in $p$ and $E$ is
	$\{ (r_1, r_2) \ | \ \textsf{head}(r_2) \in \textsf{body}(r_1)  \}$.
	Here, $\textsf{head}(r_2)$ indicates the intensional relation of the rule $r_2$, while $\textsf{body}(r_1)$ is the set of the relations in the body of the rule $r_1$. Particularly, the rule $r \in V$ is a \emph{leaf rule} if its outdegree is 0.
\end{definition}

\begin{example}
	\label{ex:rdg}
	Consider the RDG of the analysis policy in Figure~\ref{fig:input}(a).
	The vertex set $V = \{ r_1, \ r_2, \ r_3, \ r_4, \ r_5\}$ where $r_i$ indicate the rule (i) in Figure~\ref{fig:input}(a) ($ 1 \leq i \leq 5$).
	The edge set is 
	\begin{equation*}
		\begin{aligned}
			E = \{ & (r_1, r_2), (r_1, r_3), (r_1, r_4), (r_1, r_5), (r_4, r_2), (r_5, r_2), (r_4, r_3), (r_5, r_3), (r_4, r_5), (r_5, r_4), (r_4, r_4), (r_5, r_5) \}
		\end{aligned}
	\end{equation*}
    Here, $r_2$ and $r_3$ are leaf rules. $r_1$ derives the output relation $\textsf{Slice}$.
\end{example}

\begin{wrapfigure}[17]{r}{0.45\textwidth}
\vspace{-4mm}
\centering
\scalebox{1.0}{
\begin{minipage}{0.45\textwidth}
\begin{algorithm}[H]
\footnotesize
\caption{Evaluation Procedure}
\label{alg:worklist_evaluation}
\KwIn{$P$: A program; \ $p$: An analysis policy; \\
\ \ \ \ \ \ \ \ \ \ \ \emph{\textsf{Specs}}: Neural relation specifications\;}
\KwOut{$T_{R^{*}}$: The content of output relation\;}
\SetKwFunction{initializeWorklist}{InitializeWorklist}
\SetKwFunction{applyRule}{ApplyRule}
\SetKwFunction{checkWorklist}{CheckWorklist}
\SetKwFunction{updateRelations}{UpdateRelations}
\SetKwFunction{pop}{Pop}
\SetKwFunction{getHead}{GetHead}
\SetKwFunction{getLeafRules}{GetLeafRules}
\SetKwFunction{getOutputRelation}{GetOutputRelation}
\SetKwFunction{constructRuleDepGraph}{ConstructRuleDepGraph}
\SetKwFunction{reinsertDependentRules}{ReinsertDependentRules}
\SetKwFunction{getOutputRelation}{GetOutputRelation}
$(V, E) \leftarrow \constructRuleDepGraph(p)$\;
$\mathbf{S} \leftarrow [  R \mapsto \emptyset  \ | \ R \in \textsf{relations}(p) ]$\;
$W \leftarrow \getLeafRules(V, E)$\;

\While{$W$ is not empty}{
$r \leftarrow \pop(W)$; $\mathbf{S}' \leftarrow \llbracket r \rrbracket_{\emph{\textsf{Specs}}} (P, \mathbf{S})$\;
$R_I \leftarrow \getHead(r)$\;
\If{$\mathbf{S}'(R_I) \nsubseteq \mathbf{S}(R_I)$}{
\ForEach{$(r', r) \in E$}{
	$W \leftarrow W \cup \{ r' \}$\;
}
}
$\mathbf{S} \leftarrow \mathbf{S}'$\;
}
$R^{*} \leftarrow \getOutputRelation(p)$; $T_{R^{*}} \leftarrow \mathbf{S}(R^{*})$\;
\Return{$T_{R^{*}}$}\;
\end{algorithm}
\end{minipage}
}
\end{wrapfigure}
By instantiating a worklist algorithm, we present the evaluation procedure for an analysis policy on the left side of Algorithm~\hyperref[alg:worklist_evaluation]{1},
which sequentially computes the output relation in the fixed point.
Initially, it constructs an RDG $(V, E)$ at line 1 and enforces the analysis state map of each relation symbol to an empty set at line 2.
The worklist $W$ is populated with all the leaf rules at line 3.
The loop from lines 4 and 10 pops one rule $r$ from $W$ and obtains a new analysis state by computing its semantics at line 5.
If the derived relation, which is the head of the rule $r$, contains more tuples than before,
the fixed point does not reach.
In such a case,
we collect all the rules depending on the rule $r$ according to the RDG and append them to $W$.
When $W$ is empty, the algorithm terminates and reaches a fixed point.
Finally, $\mathbf{S}(R^{*})$ depicts the program property targeted in the original analysis problem, where $R^{*}$ is the output relation of the analysis policy.

\smallskip
Although Algorithm~\hyperref[alg:worklist_evaluation]{1} outlines the overall evaluation procedure for a user-specified analysis policy,
we need to concretize the rule semantics by determining how to populate the symbolic and neural relations,
which can affect the quality of the analysis result.
Meanwhile, reducing the cost of computation 
%resources and improving efficiency are also necessary for the evaluation procedure.
is important, too.
In the next few sections,
we will present the detailed technical designs that facilitate effective and efficient neuro-symbolic static analysis.

\subsection{Symbolic and Neural Constructors}
\label{subsec:rp}

\begin{definition}(Symbolic Constructor)
Given a symbolic relation symbol $R_S$,
its symbolic constructor $\gamma_{R_S}$ maps a program $P$ to a universal set of tuples belonging to $R_S$.
\end{definition}

In Table~\ref{tab:symbolic-relation-syntax}, the symbolic constructors of \textsf{Args}, \textsf{Outs}, \textsf{Paras}, \textsf{Rets}, \textsf{ExprName}, and \textsf{ExprLoc} populate these relations by inspecting the attributes of AST nodes. The constructors of \textsf{CtrlOrd} and \textsf{CtrlDep} examine the branches and loops. The constructors of \textsf{ArgPara} and \textsf{OutRet} match arguments/outputs with parameters/return values according to function names and type signatures.

\smallskip
Unlike symbolic relations, populating neural relations is more complex.
Multiple neural relation specifications may exist for a given neural relation, determining different ways of populating the relation.
Consider \textsf{DataDep} in Fig.~\ref{fig:input}(a) as an example.
Apart from Fig.~\ref{fig:input}(a), we can populate \textsf{DataDep} by prompting to check if any pair of expressions have a data dependency.
Also, we can enumerate each expression \text{e} in the program and prompt the LLMs to obtain all expressions $\text{e}'$ 
such that either $(\text{e}, \text{e}')$ or $(\text{e}', \text{e})$ belongs to \textsf{DataDep}.
Formally, we introduce the \emph{neural constructor}.

\begin{definition}(Neural Constructor)
	\label{def:nrc}
	Given a neural relation specification and an LLM,
	the induced constructor $\gamma_{R_N}$ of a binary neural relation $R_N$ can be:
	\begin{itemize}[leftmargin=*]
		\item \textbf{0-arity constructor} $\gamma_{R_N}^{0}$ maps a program $P$ to a set of pairs as the neural relation.
		\item \textbf{1-arity constructor} maps a program $P$ and the value of $x \in \textit{Dom}$ to a set of pairs that belong to $R_N$. Concretely, $\gamma_{R_N}^{1b}$ is a \emph{backward 1-arity constructor} such that $(x_1, x_2) \in \gamma_{R_N}^{1b}(P, x)$ implies $x_2$ = $x$. $\gamma_{R_N}^{1f}$ is a \emph{forward 1-arity constructor} such that $(x_1, x_2) \in \gamma_{R_N}^{1f}(P, x)$ implies $x_1$ = $x$.
		\item \textbf{2-arity constructor} $\gamma_{R_N}^{2}$ maps a program $P$ and $(x_1, x_2) \in \textit{Dom} \times \textit{Dom}$ to $\{ (x_1, x_2) \}$ or $\emptyset$.
        %indicating that $(x_1, x_2)$ belongs to or does not belong to $R_N$, respectively.
	\end{itemize}
	Similarly, the neural constructor of a unary neural relation $R_N$ is either (1) the 0-arity constructor $\gamma_{R_N}^{0}$ mapping a program $P$ to all the 1-tuples $(x) \in R_N$ or (2) the 1-arity constructor $\gamma_{R_N}^{1}$ mapping a program $P$ and $x \in \textit{Dom}$ to $\{(x)\}$ or $\emptyset$, indicating $(x)$ belongs to or does not belong to $R_N$, respectively.~\looseness=-1
\end{definition}

\begin{table}[t]
	\centering
	\caption{The examples of neural constructors and the definitions in neural relation specifications}
	\footnotesize
    \resizebox{\linewidth}{!} 
	{
	\begin{tabular}{p{0.23\linewidth} p{0.77\linewidth}}
		\toprule
		\textbf{Constructor} & \textbf{Definition} \\ \midrule
	   $\gamma_{\textsf{DataDep}}^{0}(P)$ & List $(\text{e1}, \text{e2})$ where \text{e2} is data-dependent to \text{e1}  \\
	   $\gamma_{\textsf{DataDep}}^{1b}(P, \text{e2})$ & Given \text{e2}, list $(\text{e1}, \text{e2})$ where \text{e2} is data-dependent to \text{e1}\\
	   $\gamma_{\textsf{DataDep}}^{1f}(P, \text{e1})$ &  Given \text{e1}, list $(\text{e1}, \text{e2})$ where \text{e2} is data-dependent to \text{e1} \\
	   $\gamma_{\textsf{DataDep}}^{2}(P, (\text{e1}, \text{e2}))$  &  Given \text{e1} and \text{e2}, list (\text{e1},  \text{e2}) if \text{e2} is data-dependent on \text{e1}. 
    %Otherwise, return an empty set. 
    \\ \midrule
	   $\gamma_{\textsf{XSSSrcNeural}}^{0}(P)$ & List the sources of the XSS bugs. \\
	   $\gamma_{\textsf{XSSSrcNeural}}^{1}(P, \text{e})$ & Given \text{e}, list \text{e} if \text{e} is a source of the XSS bug. 
    %Otherwise, return an empty set.  
    \\ \bottomrule
	\end{tabular}
 }
	\label{tab:neural_constructor_ex}
\end{table}

\begin{example}
\label{ex:neural_constructor}
Table~\ref{tab:neural_constructor_ex} shows the neural constructors of \textsf{DataDep} and \textsf{XSSSrcNeural}.
We also offer several examples in the neural relation specifications inducing neural constructors.
\end{example}

Although different neural constructors can all populate neural relations,
their hallucinations can vary significantly.
Intuitively, the 2-arity constructor can introduce fewer hallucinations than the 1-arity constructor when populating a binary neural relation,
as the former only focuses on the relationship between two given values.
Similarly, fewer hallucinations would be introduced by 1-arity constructors than 0-arity constructors for both unary and binary neural relations.

\subsection{Rule Evaluation}
\label{subsection:rule_evaluation}
Rule evaluation becomes complex when a neural relation is involved in the body of the rule. 
First, the LLM hallucinations introduced by different neural constructors can vary significantly.
Second, computing a fixed point in Algorithm~\ref{alg:worklist_evaluation} may require multiple rounds of prompting, consuming substantial computation resources.
In what follows, we present the technical designs by addressing the above two challenges.
Specifically, we first introduce the concept of \emph{constrained neural constructor}, which formalizes how to choose a neural constructor to populate a neural relation.
Then we demonstrate the rule evaluation using constrained neural constructors with two important strategies, namely lazy prompting and incremental prompting.~\looseness=-1

\subsubsection{Constrained Neural Constructor}
\label{subsubsection:cnc}

As discussed in Section~\ref{subsec:rp}, neural constructors with larger arities tend to reduce hallucinations when populating neural relations. 
Notably, non-neural relations can be deterministically populated, thereby constraining the values of the terms in the neural relation. 
Such terms, which we refer to as \emph{bounded terms}, enable us to choose neural constructors with as many arities as possible to effectively mitigate hallucinations.
To formalize this idea, we introduce the concept of \emph{constrained neural constructor}.

\begin{definition}(Constrained Neural Constructor)
\label{def:cnrc}
Given a set of bounded terms \( \mathbf{A} \), for a unary neural relation symbol  \( R_1 \) with the term $a$ and a binary neural relation symbol \( R_2 \) with two terms \( a_1 \) and \( a_2 \),
their constrained neural constructors are
	\[
	\footnotesize
	\begin{aligned}
		\tau(R_1, \mathbf{A}) =
		\begin{cases}
			\gamma_{R_1}^{1}, &  a \in \mathbf{A} \\
			\gamma_{R_1}^{0}, &  a \notin \mathbf{A}
		\end{cases}
		\quad
		\tau(R_2, \mathbf{A}) =
		\begin{cases}
			\gamma_{R_2}^{2}, &  a_1 \in \mathbf{A} \land a_2 \in \mathbf{A} \\
			\gamma_{R_2}^{1f}, &  a_1 \in \mathbf{A} \land a_2 \notin \mathbf{A} \\
			\gamma_{R_2}^{1b}, & a_1 \notin \mathbf{A} \land a_2 \in \mathbf{A} \\
			\gamma_{R_2}^{0}, &  a_1 \notin \mathbf{A} \land a_2 \notin \mathbf{A}
		\end{cases}
	\end{aligned}
	\]
	Here, the neural constructors $\gamma_{R_1}^{0}$, $\gamma_{R_1}^{1}$, $\gamma_{R_2}^{0}$, $\gamma_{R_2}^{1f}$, $\gamma_{R_2}^{1b}$, and $\gamma_{R_2}^{2}$ are defined in Definition~\ref{def:nrc}.
\end{definition}

\smallskip
A neural constructor searches tuples of the neural relation via prompting.
When a neural constructor has fewer arities, the search problem would be more manageable.
By enumerating the values of bounded terms and applying constrained neural constructors in Definition~\ref{def:cnrc},
we can reduce the problem of populating neural relations to a series of simpler search problems,
which can effectively reduce hallucinations.

\begin{example}
\label{ex:cnrc}
Consider the rule (4) in Fig.~\ref{fig:input}(a).
If we populate \textsf{DataDep} after obtaining the tuples in \textsf{SliceExpr}, the terms $\text{e1}$ and $\text{e2}$ become bounded, i.e., $\mathbf{A} = \{ \text{e1}, \ \text{e2}  \}$.
Hence, the constrained neural constructor of \textsf{DataDep} is the backward 1-arity constructor $\gamma_{\textsf{DataDep}}^{1b}$ as $\text{e2} \in \mathbf{A}$ and $\text{e3} \notin \mathbf{A}$.
It searches the possible value of $\text{e3}$ according to each fixed value of $\text{e2}$.
Compared to the 0-arity constructor $\gamma_{\textsf{DataDep}}^{0}$ that searches the expression pairs with data dependencies,
$\gamma_{\textsf{DataDep}}^{1b}$ concentrates on a simpler problem with reduced hallucinations.
\end{example}

\subsubsection{Rule Evaluation with Lazy Prompting}
\label{subsubsection:lp}

\input{evaluate_rule.tex}

Based on constrained neural constructors, 
we instantiate the semantics of a Datalog rule $r$ with four inference rules in Fig.~\ref{fig:semantics}.
Our basic idea is to populate the neural relation with the constrained neural constructor after joining the tuples of other relations in the rule body.
This strategy, which we call as \emph{\textbf{lazy prompting}}, can obtain the bounded terms as many as possible so that the constrained neural constructor introduces few hallucinations.
Specifically, we introduce a program $P$, an analysis state $\mathbf{S}$, a set of relations $\mathbf{R}$, and a set of bounded terms $\mathbf{A}$ as the evaluation context.
Initially, $\mathbf{R}$ contains all the relation symbols in the rule body, while $\mathbf{A}$ is empty.
By applying the four inference rules, we eventually obtain an analysis state $\llbracket r \rrbracket_{\emph{\textit{Specs}}} (P, \mathbf{S})$,
where the neural relation specifications in \emph{\textit{Specs}} instantiate all the neural constructors.
For simplification, we assume that $R_n$ is a neural relation if a rule $r := R_I \leftarrow R_1, \cdots, R_n$ contains a neural relation in its body.
Technically, the inference rules work as follows:
\begin{itemize}[leftmargin=*]
\item \textsc{Eval-Symoblic} picks a symbolic relation $R_i$ from $\mathbf{R}$, applies its symbolic constructor $\gamma_{R_i}$, and updates $\mathbf{S}$ if the symbolic relation $R_i$ has not been populated.
\textsc{Eval-Intensional} does not have a special effect on the analysis state.
The two inference rules both remove the populuated relation $R_i$ from $\mathbf{R}$
and aggregate the terms of populated relations in the set $\mathbf{A}$ as bounded ones.
	
\item If $R_n$ is a symbolic relation, \textsc{Eval-Head} conducts \emph{natural join} upon all the relations in the body and \emph{projects} the result on the terms of the intensional relation $R_I$. Here, the natural join operator $\bowtie$ and the projection operator $\Pi$ are the standard ones in relational algebra.
	
\item If $R_n$ is a neural relation,
we first apply \textsc{Eval-Symbolic} and \textsc{Eval-Intensional} until only $R_n$ is not populated.
Then \textsc{Eval-Neural} conducts the natural join on the non-neural relations and further projects the joined result upon the bounded terms in $R_n$, i.e., $\textit{term}(R_n) \cap \mathbf{A}$, which yields a set of tuples $T$.
Lastly, we apply the constrained neural constructor $\tau(R_n, \mathbf{A})$ to the tuples in $T$ as any other tuples would make $\mathbf{S}(R_1) \bowtie \cdots \mathbf{S}(R_{n-1}) \bowtie \mathbf{S}(R_{n})$ empty.
\end{itemize}

Intuitively, lazy prompting populates the neural relations until the contents of symbolic and intensional relations are determined.
It can yield as many bounded terms as possible, which enables us to choose the neural constructors with as many arities as possible,
thereby mitigating the hallucinations.
Meanwhile, the set $T$ in the inference rule \textsc{Eval-Neural} will contain more tuples if we omit several non-neural relations in the natural join, making the inference rule \textsc{Eval-Neural} induce more prompting rounds than the current design.
Hence, lazy prompting not only mitigates the hallucinations but also effectively reduces the prompting rounds in the rule evaluation.

\begin{example}
We have $\mathbf{A} = \{ \text{e1}, \ \text{e2} \}$ for \textsf{TaintProp} when evaluating the rule (1) in Fig.~\ref{fig:xss-rules}.
Using \textsc{Eval-Neural},
we populate \textsf{TaintProp} by applying $\gamma_{\textsf{TaintProp}}^{2}$ to the expressions in \textsf{XSSSrcNeural} and \textsf{XSSSinkNeural} pairwisely.
Lastly, \textsf{TaintProp} contribute to forming the output relation \textsf{XSS}.
\end{example}

\subsubsection{Rule Evaluation with Incremental Prompting}
\label{subsubsection:ip}

The inference rule \textsc{Eval-Neural} in Fig.~\ref{fig:semantics} applies a constrained neural constructor to the tuples in the set $T$. Since a Datalog rule in the analysis policy may be evaluated multiple times in the worklist, the constrained neural constructor can be repeatedly applied to the same tuple $\mathbf{t} \in T$ across different iterations. To avoid redundant prompting, we introduce \emph{\textbf{incremental prompting}} in the rule evaluation such that the constrained neural constructor is applied to the same tuple only one time.

\input{advance_neural_eval.tex}

Fig.~\ref{fig:adavanced_rule} shows the inference rule \textsc{Eval-Neural-Incremental} that improves the inference rule \textsc{Eval-Neural} in Fig.~\ref{fig:semantics} with incremental prompting.
The main difference lies in the construction of $\mathbf{S}'(R_n)$ highlighted in red.
It first projects the tuples currently existing in the relation $R_n$, i.e., $\mathbf{S}(R_n)$, to their bounded terms, i.e., the terms in $\textit{term}(R_n) \cap \mathbf{A}$,
forming the set $T_0$.
Notably, the tuples in $T_0$ are exactly the ones that have been fed to the constrained neural constructor $\widehat{\alpha}$ in previous iterations.
Hence, we only apply $\widehat{\alpha}$ upon the tuples in the difference set of $T$ and $T_0$ and directly reuse the original tuples in $\mathbf{S}(R_n)$.
We also need to memorize all the tuples $\mathbf{t}$ that make $\widehat{\alpha}(P, \mathbf{t})$ empty so that each tuple is only fed to the constrained neural constructor one time. 
By replacing \textsc{Eval-Neural} in Fig.~\ref{fig:semantics} with \textsc{Eval-Neural-Incremental}, we can avoid redundant prompting and then reduce the resource consumption during the rule evaluation.

\begin{example}
Consider evaluating the rule (4) after the rule (3) in Fig.~\ref{fig:input}(a).
After evaluating the rule (3), $\textsf{SliceExpr}$ contains $(\textsf{userCity}_{26}, \textsf{response}_{25})$ and \textsf{DataDep} is empty.
For the rule (4), the backward 1-arity constructor $\gamma_{\textsf{DataDep}}^{1b}$ generates $\textsf{response}_{23}$ as the data dependency of $\textsf{response}_{25}$,
yielding $(\textsf{userCity}_{26}, \textsf{response}_{23}) \in \text{SliceExpr}$ and $(\textsf{response}_{23}, \textsf{response}_{25}) \in \textsf{DataDep}$.
As $\textsf{response}_{25}$ has been fed to $\gamma_{\textsf{DataDep}}^{1b}$, it is unnecessary to redundantly discover the tuple $(\textsf{response}_{23}, \textsf{response}_{25})$
Hence, we only apply $\gamma_{\textsf{DataDep}}^{1b}$ to $\textsf{response}_{23}$ when we evaluate the rule (4) in another iteration.
\end{example}

%% file: evaluate_rule.tex
\begin{figure*}
\vspace{-1mm}
	\footnotesize
    \resizebox{0.92\textwidth}{!}{%
	\centering
	\begin{subfigure}[b]{0.40\linewidth}
		\centering
		\[
		\inference
		{
			r := R_I \leftarrow R_1, \cdots, R_n\\
             R_i \in \mathbf{R}, \ \ \textsf{type}(R_i) = \textsf{symbolic}, \ \ \mathbf{S}'  = \mathbf{S} \\
			\mathbf{S}' = \mathbf{S}' [R_i \mapsto \textbf{ite}(\mathbf{S}'(R_i) = \emptyset, \ \gamma_{R_i}(P), \ \mathbf{S}'(R_i))] \\
			\mathbf{R}' =  \mathbf{R_i} \setminus \{ R_i \}, \ \  \mathbf{A}' = \mathbf{A} \cup \textsf{term}(R_i)
		}
		{
			P, \ \mathbf{S}, \ \mathbf{R}, \ \mathbf{A} \vdash r \rightsquigarrow \mathbf{S}', \ \mathbf{R}', \ \mathbf{A}'
		}
		\]
		\quad (\textsc{Eval-Symbolic})

        \vspace{-2mm}
  
		\[
		\inference
		{
			r := R_I \leftarrow R_1, \cdots, R_n\\
			R_i \in \mathbf{R}, \ \ \textsf{type}(R_i) = \textsf{intensional}, \ \ \mathbf{S}'  = \mathbf{S}\\
			\mathbf{R}' =  \mathbf{R} \setminus \{ R_i \}, \ \  \mathbf{A}' = \mathbf{A} \cup \textsf{term}(R_i)
		}
		{
			P, \ \mathbf{S}, \ \mathbf{R}, \  \mathbf{A} \vdash r \rightsquigarrow \mathbf{S}', \ \mathbf{R}', \ \mathbf{A}'
		}
		\]
		\quad (\textsc{Eval-Intensional})
	\end{subfigure}
	\hfill
	\begin{subfigure}[b]{0.59\linewidth}
		\centering
		\[
		\inference
		{
			r := R_I \leftarrow R_1, \cdots, R_n\\
			R_n \in \mathbf{R}, \ \ |\mathbf{R}| = 1, \ \ \textsf{type}(R_n) = \textsf{neural}, \ \ \mathbf{S}'  = \mathbf{S} \\
			 T = \Pi_{\textsf{term}(R_n) \cap \mathbf{A}} (\mathbf{S}(R_1) \bowtie \cdots \mathbf{S}(R_{n-1}))\\
			\widehat{\alpha} = \tau(R_n, \mathbf{A}), \ \ \mathbf{S}' = \mathbf{S}' [R_n \mapsto \bigcup_{\mathbf{t} \in T} \widehat{\alpha}(P, \ \mathbf{t})]\\ \mathbf{R}' =  \mathbf{R} \setminus \{ R_n \}, \ \  \mathbf{A}' = \mathbf{A} \cup \textsf{term}(R_n)
		}
		{
			P, \ \mathbf{S}, \ \mathbf{R}, \ \mathbf{A} \vdash r \rightsquigarrow \mathbf{S}', \ \mathbf{R}', \ \mathbf{A}'
		}
		\]
		\quad (\textsc{Eval-Neural})

        \vspace{-2mm}
  
		\[
		\inference
		{
 			r := R_I \leftarrow R_1, \cdots, R_n, \ \ \mathbf{R} = \emptyset, \ \ \mathbf{S}'  = \mathbf{S}\\
 			\mathbf{S}' = \mathbf{S}'[R_I \mapsto \Pi_{\textsf{term}(R_I)} (\mathbf{S}_1(R_1)  \cdots \bowtie \mathbf{S}_1(R_n))]
		}
		{
			P, \ \mathbf{S}, \ \mathbf{R}, \  \mathbf{A} \vdash r \rightsquigarrow \mathbf{S}', \ \mathbf{R}', \ \mathbf{A}'
		}
		\]
		\quad (\textsc{Eval-Head})
	\end{subfigure}
    }
	\caption{The instantiation of rule semantics}
    \vspace{-6mm}
	\label{fig:semantics}
\end{figure*}

%% file: advance_neural_eval.tex
\begin{figure}
	\footnotesize
	\centering
		\centering
		\[
		\inference
		{
			r := R_I \leftarrow R_1, \cdots, R_n, \ \ R_n \in \mathbf{R}, \ \ |\mathbf{R}| = 1, \ \ \text{type}(R_n) = \text{neural},\ \ \mathbf{S}'  = \mathbf{S},\\
            T = \Pi_{\text{term}(R_n) \cap \mathbf{A}} (\mathbf{S}(R_1) \bowtie \cdots \mathbf{S}(R_{n-1})), \ \ 
		\widehat{\alpha} = \tau(R_n, \mathbf{A}), \ \mathbf{R}' =  \mathbf{R} \setminus \{ R_n \}, \ \  \mathbf{A}' = \mathbf{A} \cup \text{term}(R_n) \\
        \color{red}{T_0 = \Pi_{\text{term}(R_n) \cap \mathbf{A}} (\mathbf{S}(R_n))}, \ 
        \color{red}{\mathbf{S}' = \mathbf{S}' [R_n \mapsto {\color{red}{\mathbf{S}(R_n) \cup \bigcup_{\mathbf{t} \in T \setminus T_0} \widehat{\alpha}(P, \ \mathbf{t})}}]} \\
            }
		{
			P, \ \mathbf{S}, \ \mathbf{R}, \ \mathbf{A} \vdash r \rightsquigarrow \mathbf{S}', \ \mathbf{R}', \ \mathbf{A}'
		}
		\]
		\quad (\textsc{Eval-Neural-Incremental})
	\caption{The improved inference rule for a neural relation}
    \vspace{-7mm}
	\label{fig:adavanced_rule}
\end{figure}

%% file: evaluation.tex
\section{Implementation and Evaluation}
\label{sec:evaluation}

In the implementation of \ToolName,
we utilize \texttt{tree-sitter} library~\cite{brunsfeld2018tree} to implement symbolic constructors and power neural constructors with \text{GPT-3.5-Turbo}.
We set the temperature to 0 to minimize the randomness.
To evaluate \ToolName, we investigate five research questions:

\begin{itemize}[leftmargin=*]
	\item \textbf{RQ1.} \blue{How effectively does \ToolName\ support downstream analysis clients?}
    \item \textbf{RQ2.} \blue{How efficiently does \ToolName\ support downstream analysis clients?}
	\item \textbf{RQ3.} \blue{How does \ToolName\ compare against existing approaches?}
	\item \textbf{RQ4.} How do lazy prompting and incremental prompting contribute to the performance?
	\item \textbf{RQ5.} How does \ToolName\ perform in real-world bug detection?
\end{itemize}

\subsection{Clients and Datasets}	
\ \ \ \ \emph{\textbf{Program Slicing.}} To the best of our knowledge, no existing benchmarks specifically target program slicing. Although \textsc{NS-Slicer} extracts slices from CodeNet dataset~\cite{DBLP:journals/corr/abs-2105-12655} using the static slicer \textsc{JavaSlicer}~\cite{DBLP:conf/sefm/GalindoPS22}, the resulting dataset is in low quality as the slices do not account for program lines after slicing seeds. To address this, we reconfigure \textsc{JavaSlicer} and apply it to 500 programs randomly selected from the CodeNet dataset, each with specified slicing seeds, to create 500 high-quality program slices as ground truth.

\smallskip
\emph{\textbf{Bug Detection.}}
We evaluate \ToolName\ on six bug types from the Juliet Test Suite~\cite{BolandB12}, a widely used static analysis benchmark that provides detailed bug specifications and ground truth: Absolute Path Traversal (APT), Cross-Site Scripting (XSS), Divide-by-Zero (DBZ), \blue{Null Pointer Dereference (NPD), Resource Exhaustion (RE), and Unsalted One-Way Hash (UOWH).}
These bug types can be formulated as instances of data-flow analysis problems~\cite{RepsHS95}, which span diverse analytical characteristics, including source/sink definitions and value propagation.
APT and XSS represent classic taint-style vulnerabilities, while DBZ and NPD require reasoning about the propagation of specific values (e.g., zero or null) along feasible paths.
RE further demands checking whether the amount of allocated resource is data-dependent on user-controlled inputs and unbounded by control-flow conditions, whereas UOWH focuses on cryptographic misuse that relies on API-level semantics.
From a security perspective, APT and XSS consistently appear in the Top-25 CWE list, and DBZ and NPD are representative numeric and pointer-related bugs widely studied in static analysis, respectively. Also, RE and UOWH pose performance and security risks, respectively.
Overall, these well-established CWEs provide a sound basis for evaluating \ToolName’s support for customized analysis across different bug types.
Due to the high cost of a complete evaluation (See Section~\ref{subsec:discussion}), we randomly sample 100 programs per bug type that cover different patterns of sources/sinks and value propagations and include all 17 UOWH programs in our evaluation.

\subsection{\blue{Instantiation of \ToolName}}
Starting from the problem definition, we declaratively specify an analysis policy by composing a small number of symbolic relations to capture deterministic program facts, together with a few task-specific neural relations to express semantic properties that are difficult to formalize precisely. 
Each neural relation is instantiated via a prompt template as the neural relation specification, augmented with few-shot examples that define the forms of sources and sinks as well as the intended data-flow propagation patterns.
For the bug types in the Juliet Test Suite, the definitions of sources, sinks, and associated semantic constraints are briefly documented in the benchmarks,
based on which we derive natural language descriptions and construct corresponding few-shot examples with the assistance of LLMs, making the overall customization process straightforward and easy to complete.
As shown in Table~\ref{table:instantiation}, across program slicing and six bug detection tasks, the resulting policies consist of only 5--13 Datalog rules, 3--6 symbolic relations, and at most 4 neural relations, indicating modest human effort for customization. Notably, the neural relation specifications contain fewer than 60 prompt lines on average, which demonstrates substantially lightweight customization.

Concretely, we specify the analysis policy for program slicing as depicted in Fig.~\ref{fig:input}(a). For APT and XSS detection, we introduce the neural relation \textsf{TaintProp} to track taint propagation between expressions, together with neural relations encoding the corresponding sources and sinks and symbolic relations capturing function-level program values, such as parameters, arguments, and return values. For DBZ and NPD detection, we introduce \textsf{ZeroProp} and \textsf{NullProp} to trace the propagation of zero and null values, respectively, along with additional neural relations that describe the associated sources and sinks.
UOWH detection involves cryptographic API usages whose semantics are difficult to capture precisely with handcrafted symbolic checkers, but can be effectively specified using natural language descriptions and few-shot examples that leverage LLMs’ pretraining knowledge. Finally, RE bugs impose path constraints beyond simple data-flow reachability: not only must resource allocation amounts depend on user-controlled input, but the corresponding data-dependent expressions must also remain unconstrained by bounding branch or loop conditions, which is captured by the neural relation \textsf{UnboundedDataFlow}.
All analysis policies and neural relation specifications used in the evaluation are provided in~\cite{nesa_artifact}.

\begin{table}[t]
	\centering
	\caption{The numbers of the rules, symbolic relations, neural relations, and the average numbers of the lines in the neural relation specifications for program slicing and bug detection upon the Juliet Test Suite}
	\resizebox{0.7\linewidth}{!} 
	{
		\begin{tabular}{c|ccccccc}
			\toprule
			\textbf{} 
			& \textbf{Slicing} 
			& \textbf{APT} 
			& \textbf{XSS} 
			& \textbf{DBZ} 
			& \textbf{NPD} 
			& \textbf{RE} 
			& \textbf{UOWH} \\ 
			\midrule
			\textbf{\#Rules}            
			& 5  & 12 & 12 & 13 & \blue{12} & \blue{12} & \blue{12} \\
			\textbf{(\#Symbolic, \#Neural)}        
			& (3, 1) & (6, 3) & (6, 3) & (6, 4) & \blue{(6, 3)} & \blue{(6, 3)} & \blue{(6, 3)} \\
			\textbf{\# Lines} 
			& 34 & 54 & 46 & 47 & \blue{46} & \blue{47} & \blue{30} \\
			\bottomrule
		\end{tabular}
	}
    \vspace{-8mm}
	\label{table:instantiation}
\end{table}

\subsection{(RQ1): \blue{Effectiveness of NESA}}
\label{subsec:eval_effectiveness_efficiency}

\subsubsection{Setup}
We evaluate the precision and recall of \ToolName\ to quantify its effectiveness. For program slicing, the precision and recall are computed by comparing the line numbers in the reported slices against the ground-truth slices. For bug detection tasks, precision and recall are computed by comparing reported vulnerabilities with the ground truth provided by the Juliet Test Suite.

\subsubsection{Result}
Table~\ref{table:instantiation} reports the precision and recall of \ToolName\ across program slicing and six bug detection tasks. Overall, the results demonstrate that \ToolName\ achieves consistently strong effectiveness across diverse analysis scenarios.
For program slicing, \ToolName\ attains 91.50\% precision and 84.61\% recall, corresponding to an F1 score of 0.88, indicating accurate localization of relevant program statements with limited noise. For APT detection, it achieves 94.74\% precision and 90.00\% recall, while for XSS detection, the precision and recall further increase to 98.95\% and 94.00\%, respectively. These results demonstrate that \ToolName\ can precisely capture taint propagation behaviors across different forms of sources and sinks. \blue{For DBZ detection, the precision and recall decrease to 54.62\% and 71.00\%, respectively, leading to a lower F1 score. This degradation is expected, as DBZ detection requires reasoning about path feasibility and guarding conditions, whereas the current analysis focuses on data-flow reasoning without full path-condition satisfiability checking.} Despite this limitation, \ToolName\ still maintains reasonable recall, indicating its ability to identify most true DBZ cases.
\del{Besides, the columns \textbf{In}, \textbf{Out}, and \textbf{T(s)} indicate the average costs for different tasks. Specifically, the average time cost of program slicing is 6.02 seconds. \ToolName\ achieves APT, XSS, and DBZ detection with average time costs of 13.79, 29.74, and 13.78 seconds, respectively. 
The financial costs for program slicing, APT, XSS, and DBZ detection are only \$0.005, \$0.016, \$0.051, and \$0.018, respectively. These results indicate that \ToolName\ achieves promising precision and recall without significant overhead, which showcases its practical value as a customizable and compilation-free static program analyzer.}
\blue{Beyond these three tasks, \ToolName\ also demonstrates strong effectiveness on the additional bug types. Specifically, it achieves 75.45\% precision and 83.00\% recall on NPD, 97.26\% precision and 71.00\% recall on RE, and 100\% precision and recall on UOWH. These results suggest that \ToolName\ generalizes well across bug types with substantially different source–sink semantics and semantic constraints over program paths.
Therefore, \ToolName\ can effectively support diverse static program analysis tasks with high precision and recall.}

\begin{table}[t]
	\centering
	\caption{The precision and recall of \ToolName\ (\textcircled{1}), the end-to-end few-shot CoT prompting using {GPT-3.5-Turbo} (\textcircled{2}), {GPT-4-Turbo} (\textcircled{3}), {GPT-4o-mini} (\textcircled{4}), and {Claude 3.5 Sonnet} (\textcircled{5}), and non-prompting-based SOTA (\textcircled{6}).}
	\resizebox{0.95\linewidth}{!}{
		\begin{tabular}{c|cc|cc|cc|cc|cc|cc|cc}
			\toprule
			& \multicolumn{2}{c|}{\textbf{Slicing}}
			& \multicolumn{2}{c|}{\textbf{APT}}
			& \multicolumn{2}{c|}{\textbf{XSS}}
			& \multicolumn{2}{c|}{\textbf{DBZ}}
			& \multicolumn{2}{c|}{\blue{\textbf{NPD}}}
			& \multicolumn{2}{c|}{\blue{\textbf{RE}}}
			& \multicolumn{2}{c}{\blue{\textbf{UOWH}}} \\
			
			& \textbf{P(\%)} & \textbf{R(\%)}
			& \textbf{P(\%)} & \textbf{R(\%)}
			& \textbf{P(\%)} & \textbf{R(\%)}
			& \textbf{P(\%)} & \textbf{R(\%)}
			& {\blue{\textbf{P(\%)}}} & {\blue{\textbf{R(\%)}}}
			& {\blue{\textbf{P(\%)}}} & {\blue{\textbf{R(\%)}}}
			& {\blue{\textbf{P(\%)}}} & {\blue{\textbf{R(\%)}}} \\
			\midrule
			
			\textcircled{1} 
			& 91.50 & 84.61 & 94.74 & 90.00 & 98.95 & 94.00 & 54.62 & 71.00
			& 75.45 & 83.00 & 97.26 & 71.00 & 100.00 & 100.00 \\ 
			
			\textcircled{2} 
			& 48.92 & 53.82 & 31.58 & 78.00 & 46.32 & 88.00 &  1.38 &  3.00
			& 41.04 & 71.00 & 47.79 & 65.00 & 78.95 & 88.24 \\
			
			\textcircled{3} 
			& 65.29 & 48.90 & 76.36 & 100.00 & 87.38 & 97.00 & 57.00 & 57.00
			& 58.21 & 78.00 & 61.61 & 69.00 & 89.47 & 100.00 \\
			
			\textcircled{4} 
			& 65.92 & 47.89 & 41.28 & 97.00 & 81.15 & 99.00 & 26.17 & 89.00
			& 51.22 & 84.00 & 67.03 & 61.00 & 78.95 & 88.24 \\
			
            \textcircled{5} 
			& 53.65 & 40.82 & 91.09 & 92.00 & 75.76 & 100.00 & 51.18 & 65.00
			& 60.61 & 80.00 & 65.18 & 73.00 & 100.00 & 100.00 \\ 
			
			\textcircled{6} 
			& 69.81 & 99.03 & 100.00 & 78.00 & 100.00 & 54.00 & 88.31 & 68.00
			& 78.38 & 87.00 & 43.98 & 84.00 & NA & NA \\
			
			\bottomrule
		\end{tabular}
	}
	\label{table:baseline_cmp}
    \vspace{-6mm}
\end{table}

\begin{wrapfigure}[13]{r}{0.31\textwidth}
	\centering
	\vspace{-10mm}
	\scalebox{1.0}{
		\begin{minipage}{0.3\textwidth}
			\centering
			\begin{table}[H]
            \centering
            \caption{\blue{The statistics of the input/output token and time costs per program slicing case of \ToolName\ and prompting-based baselines. Each cell $v_\sigma$ indicates the average value $v$ and the standard deviation $\sigma$~\cite{KitchenhamRobustStats}. \textcircled{1}--\textcircled{5} have the same meanings as the ones in Table~\ref{table:baseline_cmp}.}}
            \resizebox{0.95\linewidth}{!}{
            \bluetable{\begin{tabular}{lccc}
            \toprule
             & \textbf{In(K)} & \textbf{Out(K)} & \textbf{T(s)} \\
            \midrule
            \textcircled{1}   & $7.81_{0.31}$ & $0.70_{0.08}$ & $6.02_{0.72}$ \\
            \textcircled{2}   & $1.20_{0.05}$ & $0.10_{0.02}$ & $3.17_{0.55}$ \\
            \textcircled{3}   & $1.20_{0.05}$ & $0.10_{0.03}$ & $5.12_{0.98}$ \\
            \textcircled{4}   & $1.20_{0.05}$ & $0.10_{0.02}$ & $4.19_{0.70}$ \\
            \textcircled{5}   & $1.20_{0.05}$ & $0.10_{0.03}$ & $9.76_{1.98}$ \\
            \bottomrule
            \end{tabular}
            }}
            \label{tab:baseline_cost_slicing}
            \end{table}
		\end{minipage}
	}
\end{wrapfigure}
\subsection{\blue{(RQ2): Efficiency of \ToolName}}
\subsubsection{Setup}
\blue{We evaluate the efficiency of \ToolName\ on program slicing and six bug detection tasks by measuring input token cost, output token cost, financial cost, and time cost. Also, we measure the standard deviations~\cite{KitchenhamRobustStats} of the input token cost, output token cost, and time cost across different analyzed programs.
The detailed statistics are summarized in the first row of Tables~\ref{tab:baseline_cost_slicing} and~\ref{tab:cost_baseline_bug_detection}.}

\subsubsection{Result}
\blue{Overall, \ToolName\ exhibits stable and predictable computational costs across all tasks, as shown in Tables~\ref{tab:baseline_cost_slicing} and~\ref{tab:cost_baseline_bug_detection}.
For program slicing, \ToolName\ incurs an average time cost of 6.02 seconds. Based on the pricing of OpenAI API~\cite{openai2026pricing}, the average financial cost is 0.495 cents. For bug detection, the computational costs naturally increase with the complexity of analysis policies and neural relations. As shown in Table~\ref{tab:cost_baseline_bug_detection}, across different bug types, the average execution time ranges from several seconds to a few tens of seconds, while the financial cost consistently stays at the level of only a few cents per task.
The standard deviations of the input token cost are 
primarily attributed to the varying program sizes. For example, in APT bug detection, the programs have an average length of 276.87 lines with a standard deviation of 57.04 (ranging from 188 to 429 lines), which directly leads to the observed variance in token consumption and time cost across different cases.
Among these bug types, XSS detection incurs higher input/output token usage and financial cost, with average input tokens exceeding 70K and costs reaching approximately \$0.05. This is primarily because XSS bugs involve more diverse and complex forms of sources and sinks than other bug types, requiring longer neural relation specifications. Nevertheless, XSS detection still completes within roughly 30 seconds on average. In contrast, DBZ detection exhibits lower token consumption and time overhead.
For NPD, RE, and UOWH, \ToolName\ maintains similar efficiency trends. Although these tasks differ in semantic constraints and context requirements, their input and output token usage remains within the same order of magnitude (typically below 25K input tokens), and average execution times remain well under 20 seconds.}

\begin{table*}[t]
	\centering
	\caption{\blue{The statistics of the input token, output token, and time costs per bug detection case of \ToolName\ and prompting-based baselines on the Juliet Test Suite. $v_\sigma$ in each cell indicates the average value $v$ and the standard deviation $\sigma$ of the values~\cite{KitchenhamRobustStats}. \textcircled{1}--\textcircled{5} have the same meanings as the ones in Table~\ref{table:baseline_cmp}.}}
	\resizebox{0.8\textwidth}{!}{
	\bluetable{\begin{tabular}{c|ccc|ccc|ccc}
	\toprule
	& \multicolumn{3}{c|}{\textbf{APT}}
	& \multicolumn{3}{c|}{\textbf{XSS}}
	& \multicolumn{3}{c}{\textbf{DBZ}} \\
	\cmidrule(lr){2-4} \cmidrule(lr){5-7} \cmidrule(lr){8-10}
	& \textbf{In(K)} & \textbf{Out(K)} & \textbf{T(s)}
	& \textbf{In(K)} & \textbf{Out(K)} & \textbf{T(s)}
	& \textbf{In(K)} & \textbf{Out(K)} & \textbf{T(s)} \\
	\midrule
	\textcircled{1}
	& $27.40_{5.81}$ & $1.50_{0.38}$ & $13.79_{3.21}$
	& $72.20_{6.51}$ & $9.70_{1.93}$ & $29.74_{5.87}$
	& $29.10_{5.94}$ & $2.30_{0.51}$ & $13.78_{2.87}$ \\
	\textcircled{2}
	& $7.65_{1.30}$ & $0.37_{0.07}$ & $4.07_{0.70}$
	& $8.63_{1.08}$ & $0.28_{0.06}$ & $3.17_{0.59}$
	& $6.54_{1.05}$ & $0.45_{0.09}$ & $3.95_{0.72}$ \\
	\textcircled{3}
	& $7.65_{1.30}$ & $0.42_{0.08}$ & $9.12_{1.67}$
	& $8.63_{1.08}$ & $0.18_{0.04}$ & $8.10_{1.41}$
	& $6.54_{1.05}$ & $0.54_{0.11}$ & $8.47_{1.60}$ \\
	\textcircled{4}
	& $7.65_{1.30}$ & $0.33_{0.06}$ & $6.01_{1.05}$
	& $8.63_{1.08}$ & $0.30_{0.07}$ & $5.71_{1.04}$
	& $6.54_{1.05}$ & $0.47_{0.08}$ & $5.34_{0.91}$ \\
	\textcircled{5}
	& $7.65_{1.30}$ & $0.58_{0.12}$ & $10.44_{1.99}$
	& $8.63_{1.08}$ & $0.45_{0.09}$ & $11.26_{2.23}$
	& $6.54_{1.05}$ & $0.67_{0.13}$ & $9.05_{1.67}$ \\
	\midrule
	& \multicolumn{3}{c|}{\textbf{NPD}}
	& \multicolumn{3}{c|}{\textbf{RE}}
	& \multicolumn{3}{c}{\textbf{UOWH}} \\
	\cmidrule(lr){2-4} \cmidrule(lr){5-7} \cmidrule(lr){8-10}
	& \textbf{In(K)} & \textbf{Out(K)} & \textbf{T(s)}
	& \textbf{In(K)} & \textbf{Out(K)} & \textbf{T(s)}
	& \textbf{In(K)} & \textbf{Out(K)} & \textbf{T(s)} \\
	\midrule
	\textcircled{1}
	& $19.90_{4.12}$ & $2.87_{0.71}$ & $17.47_{3.54}$
	& $22.82_{4.63}$ & $2.18_{0.52}$ & $14.39_{2.98}$
	& $8.87_{2.47}$ & $0.75_{0.22}$ & $7.77_{2.01}$ \\
	\textcircled{2}
	& $5.82_{0.78}$ & $0.27_{0.05}$ & $3.07_{0.53}$
	& $5.98_{1.06}$ & $0.35_{0.07}$ & $3.26_{0.61}$
	& $5.36_{0.85}$ & $0.25_{0.05}$ & $3.95_{0.69}$ \\
	\textcircled{3}
	& $5.82_{0.78}$ & $0.32_{0.06}$ & $7.12_{1.31}$
	& $5.98_{1.06}$ & $0.44_{0.09}$ & $8.09_{1.43}$
	& $5.36_{0.85}$ & $0.27_{0.05}$ & $7.12_{1.36}$ \\
	\textcircled{4}
	& $5.82_{0.78}$ & $0.23_{0.04}$ & $5.19_{0.97}$
	& $5.98_{1.06}$ & $0.37_{0.08}$ & $7.34_{1.28}$
	& $5.36_{0.85}$ & $0.26_{0.06}$ & $6.01_{1.06}$ \\
	\textcircled{5}
	& $5.82_{0.78}$ & $0.38_{0.08}$ & $10.28_{1.97}$
	& $5.98_{1.06}$ & $0.47_{0.09}$ & $13.44_{2.76}$
	& $5.36_{0.85}$ & $0.35_{0.07}$ & $10.91_{2.12}$ \\
	\bottomrule
	\end{tabular}}}
	\label{tab:cost_baseline_bug_detection}
	\vspace{-8mm}
	\end{table*}

\subsection{(RQ3): Comparison with Existing Techniques}
\label{subsec:eval_existing_technique}

\subsubsection{Setup}

%We select two categories of baselines. The first one comprises 

We compare \ToolName\ with end-to-end few-shot CoT prompting-based methods powered by {GPT-3.5-Turbo}, {GPT-4-Turbo}, {GPT-4o-mini}, and {Claude-3.5-Sonnet}, which represent the latest in cost-efficient models within this lineage. 
We also consider non-prompting-based techniques.
In the program slicing, we compare \ToolName\ with \textsc{NS-Slicer}~\cite{DBLP:journals/pacmpl/YadavallyLWN24}, a recent learning-based slicing method configured with GraphCodeBERT. For bug detection, we compare \ToolName\ with \textsc{Pinpoint}~\cite{Shi18Pinpoint}, the SOTA bug detection tool.
It does not support non-intrusive customization and is compilation-dependent.
We measure the precision, recall, and computational costs of the baselines.

\subsubsection{Result}
As shown in Table~\ref{table:baseline_cmp}, \ToolName\ consistently outperforms few-shot CoT prompting with different LLMs in program slicing.
Based on precision and recall, the F1 score of \ToolName\ surpasses those of the four LLM-based baselines (i.e., \textcircled{2}–\textcircled{5}) by 0.37, 0.32, 0.33, and 0.42, respectively.
In contrast, \textsc{NS-Slicer}, which does not leverage LLMs for slice prediction, achieves only 69.81\% precision and an F1 score of 0.82.
For APT and XSS detection, all four LLMs achieve high recall due to the simplicity of the bug patterns, where return values and function arguments naturally act as sources and sinks.
While LLMs reliably identify these program values, they are less effective at precisely reasoning about taint propagation, leading to reduced precision.
For example, GPT-3.5-Turbo attains only 31.58\% and 46.32\% precision on APT and XSS, respectively.
For DBZ and NPD detection, end-to-end prompting-based approaches exhibit substantially lower precision, as these bug types require accurate reasoning about path conditions to exclude infeasible program paths.
Similarly, RE detection demands checking whether program values are bounded by path conditions, a form of reasoning that is difficult to achieve through end-to-end prompting and results in both low precision and recall.
In contrast, all UOWH buggy cases are intra-procedural, allowing end-to-end prompting to achieve comparatively higher precision and recall than for other bug types.
Compared with prompting-based approaches, the non-prompting-based solution \textsc{Pinpoint} exhibits a distinct trade-off.
Its restrictive built-in source and sink specifications lead to recall loss for APT and XSS detection.
In addition, insufficient modeling of library function semantics introduces false positives when reasoning about path feasibility.
Overall, \ToolName\ and \textsc{Pinpoint} achieve comparable precision and recall on APT, XSS, NPD, and DBZ.
However, for RE detection, \textsc{Pinpoint} cannot determine whether specific program values are bounded, resulting in a substantially higher false-positive rate.
Lastly, \textsc{Pinpoint} lacks semantic modeling of cryptographic libraries, so it cannot be extended to support UOWH detection, whereas \ToolName\ naturally generalizes to this bug type by leveraging prior knowledge acquired by LLMs during pre-training.

Tables~\ref{tab:baseline_cost_slicing} and~\ref{tab:cost_baseline_bug_detection} report the token usage, financial cost, and time cost of \ToolName\ and the prompting-based baselines, with per-cell standard deviations shown as subscripts.
Although \ToolName\ incurs higher token, financial, and time costs than end-to-end prompting baselines on a per-case basis, it detects substantially more true bugs.
When normalizing financial cost by recall, \ToolName\ achieves an average cost–recall ratio of 0.023 across bug types, which is significantly lower than that of GPT-4-Turbo (0.097) and Claude~3.5~Sonnet (0.033).
This indicates that \ToolName\ incurs less financial cost to detect a single true bug than the baselines using stronger models.

\subsection{(RQ4): Ablation Studies}
\subsubsection{Setup}
We introduce three ablations:
\textsc{NESA-1AB} and \textsc{NESA-1AF} utilizes backward and forward 1-arity constructors as the constrained neural constructors of binary neural relations, respectively.
\textsc{NESA-0A} utilizes 0-arity constructors for both unary and binary neural relations. 
We evaluate them with the same metrics as Section~\ref{subsec:eval_effectiveness_efficiency}.
Due to potential high token costs, we do not assess \ToolName\ without incremental prompting, and instead, quantify the skipped prompting rounds.

\begin{table*}[t]
\centering
\caption{The precision and recall of \ToolName\ (\textcircled{1}), NESA-1AB (\textcircled{2}), NESA-1AF (\textcircled{3}), and NESA-0A (\textcircled{4})}
\resizebox{0.95\linewidth}{!}{
\begin{tabular}{c|cc|cc|cc|cc|cc|cc|cc}
\toprule
& \multicolumn{2}{c|}{\textbf{Slicing}}
& \multicolumn{2}{c|}{\textbf{APT}}
& \multicolumn{2}{c|}{\textbf{XSS}}
& \multicolumn{2}{c|}{\textbf{DBZ}}
& \multicolumn{2}{c|}{\blue{\textbf{NPD}}}
& \multicolumn{2}{c|}{\blue{\textbf{RE}}}
& \multicolumn{2}{c}{\blue{\textbf{UOWH}}} \\
& \textbf{P(\%)} & \textbf{R(\%)}
& \textbf{P(\%)} & \textbf{R(\%)}
& \textbf{P(\%)} & \textbf{R(\%)}
& \textbf{P(\%)} & \textbf{R(\%)}
& \blue{\textbf{P(\%)}} & \blue{\textbf{R(\%)}}
& \blue{\textbf{P(\%)}} & \blue{\textbf{R(\%)}}
& \blue{\textbf{P(\%)}} & \blue{\textbf{R(\%)}} \\
\midrule
\textcircled{1}
& 91.50 & 84.61
& 94.74 & 90.00
& 98.95 & 94.00
& 54.62 & 71.00
& 75.45 & 83.00
& 97.26 & 71.00
& 100.00 & 100.00 \\
\textcircled{2}
& 91.50 & 84.61
& 88.89 & 16.00
& 100.00 & 20.00
& 48.08 & 25.00
& 33.49 & 73.00
& 74.58 & 44.00
& 100.00 & 94.00 \\
\textcircled{3}
& NA & NA
& 94.74 & 54.00
& 94.67 & 49.00
& 39.90 & 77.00
& 63.55 & 68.00
& 100.00 & 37.00
& 100.00 & 82.35 \\
\textcircled{4}
& 52.14 & 16.13
& 0.00 & 0.00
& 0.00 & 0.00
& 0.00 & 0.00
& 41.98 & 55.00
& 0.00 & 0.00
& 100.00 & 5.88 \\
\bottomrule
\end{tabular}}
\label{table:ablation_metrics}
\vspace{-4mm}
\end{table*}

\begin{table*}[t]
\centering
\caption{\blue{The statistics of the input token, output token, and time costs per bug detection case of \ToolName\ (\textcircled{1}), NESA-1AB (\textcircled{2}), NESA-1AF (\textcircled{3}), and NESA-0A (\textcircled{4}) upon the Juliet Test Suite. $v_\sigma$ in each cell indicates the average value $v$ and the standard deviation $\sigma$ of the values~\cite{KitchenhamRobustStats}.}}
\resizebox{0.8\textwidth}{!}{
\bluetable{\begin{tabular}{c|ccc|ccc|ccc}
\toprule
& \multicolumn{3}{c|}{\textbf{APT}}
& \multicolumn{3}{c|}{\textbf{XSS}}
& \multicolumn{3}{c}{\textbf{DBZ}} \\
\cmidrule(lr){2-4} \cmidrule(lr){5-7} \cmidrule(lr){8-10}
& \textbf{In(K)} & \textbf{Out(K)} & \textbf{T(s)}
& \textbf{In(K)} & \textbf{Out(K)} & \textbf{T(s)}
& \textbf{In(K)} & \textbf{Out(K)} & \textbf{T(s)} \\
\midrule
\textcircled{1}
& $27.40_{5.81}$ & $1.50_{0.38}$ & $13.79_{3.21}$
& $72.20_{6.51}$ & $9.70_{1.93}$ & $29.74_{5.87}$
& $29.10_{5.94}$ & $2.30_{0.51}$ & $13.78_{2.87}$ \\
\textcircled{2}
& $27.70_{6.23}$ & $1.60_{0.45}$ & $14.22_{3.17}$
& $22.30_{6.45}$ & $1.70_{0.38}$ & $12.89_{2.67}$
& $28.30_{6.34}$ & $2.90_{0.72}$ & $14.53_{3.15}$ \\
\textcircled{3}
& $27.20_{5.30}$ & $1.50_{0.33}$ & $14.48_{3.48}$
& $28.30_{6.53}$ & $2.40_{0.44}$ & $15.55_{2.94}$
& $25.90_{4.79}$ & $1.80_{0.36}$ & $10.93_{2.13}$ \\
\textcircled{4}
& $32.40_{5.67}$ & $4.10_{1.20}$ & $23.95_{5.22}$
& $19.40_{6.48}$ & $2.90_{0.70}$ & $16.96_{5.60}$
& $25.40_{6.02}$ & $2.90_{0.56}$ & $12.47_{2.78}$ \\
\midrule
& \multicolumn{3}{c|}{\textbf{NPD}}
& \multicolumn{3}{c|}{\textbf{RE}}
& \multicolumn{3}{c}{\textbf{UOWH}} \\
\cmidrule(lr){2-4} \cmidrule(lr){5-7} \cmidrule(lr){8-10}
& \textbf{In(K)} & \textbf{Out(K)} & \textbf{T(s)}
& \textbf{In(K)} & \textbf{Out(K)} & \textbf{T(s)}
& \textbf{In(K)} & \textbf{Out(K)} & \textbf{T(s)} \\
\midrule
\textcircled{1}
& $19.90_{4.12}$ & $2.87_{0.71}$ & $17.47_{3.54}$
& $22.82_{4.63}$ & $2.18_{0.52}$ & $14.39_{2.98}$
& $8.87_{2.47}$ & $0.75_{0.22}$ & $7.77_{2.01}$ \\
\textcircled{2}
& $20.51_{4.67}$ & $3.40_{0.93}$ & $16.28_{3.48}$
& $29.18_{4.72}$ & $3.01_{0.78}$ & $14.69_{2.87}$
& $8.47_{2.15}$ & $0.77_{0.20}$ & $6.90_{1.70}$ \\
\textcircled{3}
& $17.77_{3.38}$ & $3.10_{0.67}$ & $18.20_{3.49}$
& $29.98_{4.75}$ & $2.86_{0.60}$ & $15.93_{3.44}$
& $10.93_{2.53}$ & $1.34_{0.43}$ & $9.41_{2.54}$ \\
\textcircled{4}
& $15.31_{4.03}$ & $2.69_{0.75}$ & $15.34_{3.34}$
& $29.65_{4.70}$ & $3.32_{0.90}$ & $18.46_{4.43}$
& $11.23_{2.50}$ & $1.38_{0.36}$ & $16.15_{3.94}$ \\
\bottomrule
\end{tabular}}}
\label{tab:ablation_cost_bugdetection}
\vspace{-5mm}
\end{table*}

\begin{wrapfigure}[12]{r}{0.32\textwidth}
	\vspace{-10mm}
	\centering
	\scalebox{1.0}{
		\begin{minipage}{0.31\textwidth}
			\centering
			\begin{table}[H]
            \centering
            \caption{\blue{The statistics of the input/output token and time costs per program slicing case of \ToolName\ (\textcircled{1}), NESA-1AB (\textcircled{2}), NESA-1AF (\textcircled{3}), and NESA-0A (\textcircled{4}). Each cell $v_\sigma$ indicates the average value $v$ and the standard deviation $\sigma$~\cite{KitchenhamRobustStats}.}}
            \resizebox{0.9\linewidth}{!}{
            \bluetable{\begin{tabular}{lccc}
            \toprule
             & \textbf{In(K)} & \textbf{Out(K)} & \textbf{T(s)} \\
            \midrule
            \textcircled{1} & $7.81_{0.31}$ & $0.70_{0.08}$ & $6.02_{0.72}$ \\
            \textcircled{2} & $7.81_{0.31}$ & $0.70_{0.08}$ & $6.02_{0.72}$ \\
            \textcircled{3} & NA  & NA  & NA   \\
            \textcircled{4} & $1.60_{0.29}$ & $0.61_{0.07}$ & $6.74_{0.81}$ \\
            \bottomrule
            \end{tabular}}}
            \label{tab:ablation_cost_slicing}
            \end{table}
		\end{minipage}
	}
\end{wrapfigure}
\subsubsection{Result}
Table~\ref{table:ablation_metrics} demonstrates the precision, recall, and F1 score of \ToolName\ and its ablations, while Tables~\ref{tab:ablation_cost_bugdetection} and~\ref{tab:ablation_cost_slicing} show the computational costs.
In program slicing, \textsc{NESA-0A} achieves a precision of 52.14\%, a recall of 16.13\%, and a F1 score of 0.25, all significantly lower than those of \ToolName. Notably, \textsc{NESA-1AB} is exactly \ToolName\ as both use the backward 1-arity constructor to populate \textsf{DataDep}. In this case, the first term of \textsf{DataDep} remains unbounded in each rule, making \textsc{NESA-1AF} inapplicable for program slicing.
Similar trends can be observed in bug detection tasks, highlighting the advantages of constrained neural constructors in rule evaluation. \blue{However, the superiority of \ToolName\ is less significant in the DBZ and NPD detection compared to APT and XSS detection. As the DBZ detection involves determining the satisfiability of path conditions, the 2-arity neural constructor still suffers from hallucinations when determining value flows via prompting, which reduces its overall advantage. 
For RE detection, which requires determining whether program values are bounded by path conditions, \ToolName\ exhibits a clearer advantage.
For UOWH, which is intra-procedural in nature, the performance differences among variants are comparatively smaller.}
\blue{In terms of efficiency, the average prompting rounds are 7.12, 7.00, 45.50, 12.01, 23.84, 17.34, and 8.71 for program slicing, APT, XSS, DBZ, NPD, RE, and UOWH detection, respectively, while 12.52, 33.44, 19.66, 10.86, 26.73, 17.91, and 8.12 rounds are skipped, showing that incremental prompting avoids 63.74\%, 82.69\%, 30.17\%, 47.49\%, 52.86\%, 50.81\%, and 48.25\% rounds.}

\subsection{(RQ5): Utility of Analyzing Real-world Programs}

\subsubsection{Setup}

We evaluate \ToolName\ on TaintBench~\cite{LuoPPBPMBHM22}, which contains 39 real-world Android malware applications with annotated taint flows.
\blue{We target 70 vulnerabilities introduced by intra-file taint flows. The remaining cases are primarily related to event-driven behaviors, which are orthogonal to our work.}
We import the Android applications into Cursor, an AI-assisted code editor, and use different LLMs to detect taint flows.
In addition to the models evaluated in Section~\ref{subsec:eval_existing_technique}, we also include the advanced reasoning model Claude~3.7~Sonnet.
\blue{Due to compatibility issues with Gradle, we are unable to compile TaintBench and therefore adopt \textsc{CodeFuseQuery}~\cite{sparrowTool, shi2025datalog} for compilation-free analysis.}
By manually crafting queries, we enable customized analysis for different source–sink pairs.
\blue{Furthermore, we evaluate \ToolName\ on two large real-world projects, h3 (21 KLoC) and dynamips (105 KLoC), to detect memory leak bugs and demonstrate its practical impact.}

\begin{table}[t]
	\centering
	\caption{The precision and recall of \ToolName\ (\textcircled{1}), GPT-3.5-Turbo (\textcircled{2}), GPT-4-Turbo (\textcircled{3}), GPT-4o-mini (\textcircled{4}), {Claude 3.5 Sonnet} (\textcircled{5}), {Claude 3.7 Sonnet} (\textcircled{6}), and \textsc{CodeFuseQuery} (\textcircled{7}) upon TaintBench}
	\resizebox{0.6\linewidth}{!} 
	{
		\begin{tabular}{cccccccc}
			\toprule
			\textbf{Metrics} & \textcircled{1} & \textcircled{2} & \textcircled{3} & \textcircled{4} & \textcircled{5} & \textcircled{6} & \textcircled{7} \\ \midrule
			\textbf{P (\%)}      & 66.27 & 40.00 & 63.93 & 46.97 & 66.18 & 70.59 & 70.73 \\
			\textbf{R (\%)}      & 78.57 & 22.86 & 55.71 & 44.29 & 64.29 & 68.57 & 41.43 \\
			\bottomrule
		\end{tabular} 
	}
	\label{table:model_comparison}
	\vspace{-5mm}
\end{table}

\subsubsection{Result}
As shown in Table~\ref{table:model_comparison}, GPT-3.5-Turbo and GPT-4o-mini both achieve low precision and recall.
GPT-4-Turbo achieves significantly better performance, particularly with a precision of 63.93\%, which almost reaches the precision of \ToolName. 
Claude-3.5-Sonnet achieves a precision of 66.18\% and a recall of 64.29\%.
Benefiting from the powerful reasoning ability, the precision and recall of Claude-3.7-Sonnet reach 70.59\% and 68.57\%, respectively, both comparable to \ToolName.
\blue{Its false positives and negatives are primarily caused by incorrectly identified sources and sinks in the large files.}
\textsc{\textsc{CodeFuseQuery}} achieves a precision of 70.73\% and a recall of 41.43\%. While it does not exhibit hallucinations, several false positives arise from its context and flow insensitivities. Its lack of support for analyzing specific program constructs, such as Java collections, also hinders its ability to detect taint flows in these cases and thus cause false negatives. 
\blue{As for efficiency, \ToolName\ completes the analysis of all Android malware applications in 1.98 hours at a cost of \$3.52, whereas \textsc{CodeFuseQuery} requires a total of 3.28 hours.
Because sources and sinks are relatively rare in these applications, \ToolName\ does not need to reason about data-flow facts for all functions.
In contrast, the preprocessing stage of \textsc{CodeFuseQuery} computes many unnecessary program facts that are ultimately not used during query evaluation, which make it not as efficient as \ToolName.}

\begin{wrapfigure}[12]{r}{0.47\textwidth}
\vspace{-5mm}
\centering
\scalebox{1.0}{
\begin{minipage}{0.47\textwidth}
\centering
\includegraphics[width=\linewidth]{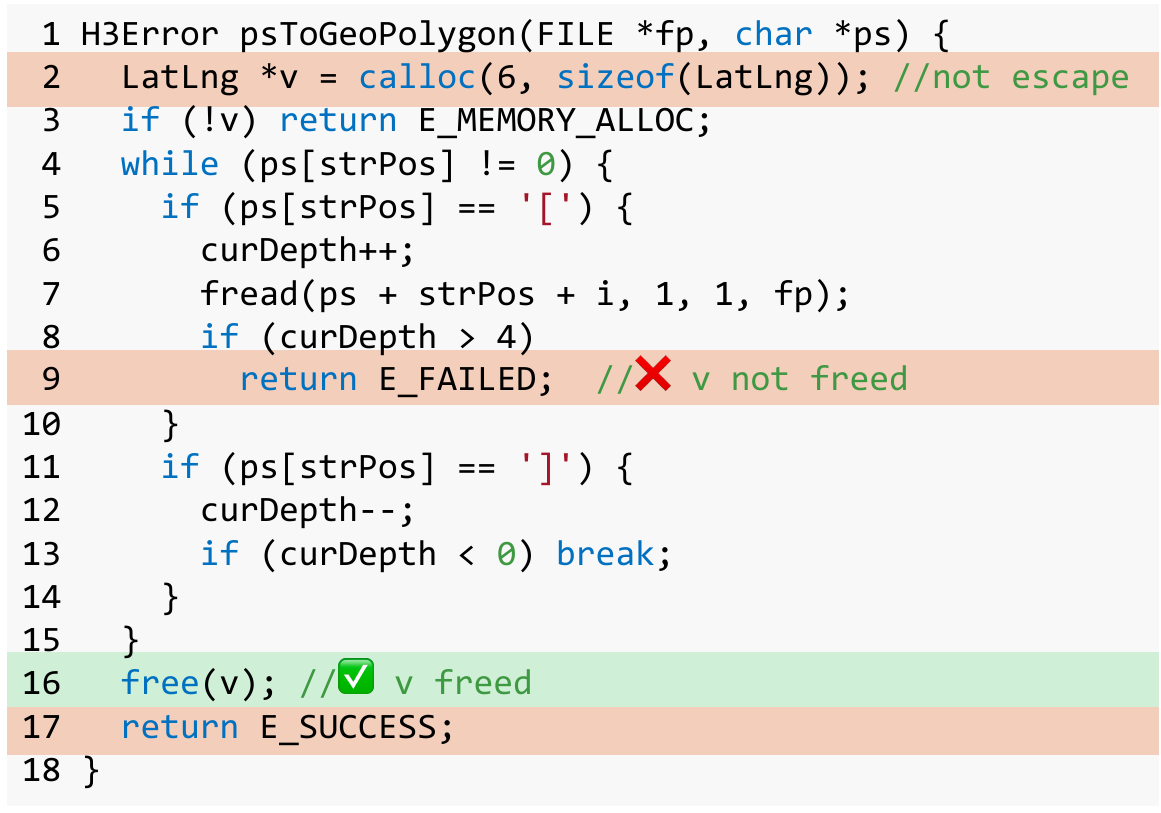}
\caption{A simplified memory leak in Uber h3}
\label{fig:case_study}
\end{minipage}
}
\end{wrapfigure}
In the projects h3 and dynamips, \ToolName{} detects 18 memory leaks, 13 of which are true bugs, yielding a precision of 72.22\%.
Fig.~\ref{fig:case_study} shows a memory leak in h3, used in Uber's production line.
As shown in the supplementary material, the symbolic relation \emph{Outs} and neural relation \emph{AllocObj} identify the pointer \texttt{verts} pointing to the object allocated at line~2.
Based on neural relation specifications of \emph{notEscape} and \emph{notFree}, the LLM determines that the object pointed by \texttt{verts} is
confined within the function and not freed before the return at line~9, causing the memory leak.
Cursor, powered by Claude 3.7 Sonnet, fails to detect any true bugs in the two projects, underscoring the superiority of neuro-symbolic design in \ToolName.
We provide developers with bug reports that include related memory allocation sites and step-by-step logs identifying relevant functions. All 13 confirmed bugs were subsequently fixed~\cite{NESABugList}.
Lastly, analyzing h3 and dynamips takes 0.09 and 0.72 hours, respectively, incurring total costs of \$0.36 and \$2.32.

\subsection{Discussion}
\label{subsec:discussion}
\ \ \ \ \emph{\textbf{Threats to Validity.}}
\emph{First}, the inherent stochasticity of LLMs may threaten reproducibility and consistency.
In our evaluation, we mitigate randomness by setting the temperature to 0 and executing \ToolName, and its baselines and ablations five times. All the statistics in the evaluation section are the average of the five runs.
\emph{Second}, the choice of the underlying LLM may affect the performance of \ToolName.
Due to budget constraints, we use GPT-3.5-Turbo as the default model in our experiments.
Nevertheless, we additionally evaluate \ToolName\ using GPT-4-Turbo on the Juliet Test Suite across all six bug types.
Besides DBZ detection, where GPT-4-Turbo achieves 91.41\% precision and 97.00\% recall, and NPD detection with 97.00\% precision and 97.00\% recall, \ToolName\ attains 100\% precision and recall on all remaining bug types.
These results confirm that \ToolName\ can seamlessly benefit from stronger reasoning capabilities provided by more advanced LLMs.

\ \ \ \emph{\textbf{Total Financial Cost and Its Breakdown.}}
To assess experimental stability, we repeat all benchmark experiments five times for \ToolName\ powered by GPT-3.5-Turbo, baselines, and ablations.
According to the pricing of OpenAI API~\cite{openai2026pricing}, conducting a single complete evaluation incurs a total cost of \$705.92. Specifically, program slicing accounts for \$62.82. Bug detection on the Juliet Test Suite accounts for \$477.76 in total, including \$304.52 for APT/XSS/DBZ and \$173.24 for NPD/RE/UOWH. Real-world program evaluation accounts for \$165.34 in total. This includes \$30.52 for \ToolName\ on TaintBench and \$132.14 for LLM-based baselines on TaintBench. For h3 and dynamips, \ToolName\ incurs \$0.36 and \$2.32, respectively. The baseline evaluation on these two projects is conducted via Cursor, whose exact cost cannot be directly measured.
We also conduct an exploratory experiment of \ToolName\ driven by GPT-4-Turbo on the Juliet Test Suite (run once) to demonstrate that improvements in the underlying LLM can seamlessly improve \ToolName's precision and recall, costing \$227.70 in total. Due to the high total cost of evaluating \ToolName, its baselines, and ablations, we do not scale experiments to the full size of benchmarks.

\smallskip
\blue{\emph{\textbf{Stability of Results.}}
We also measure the stability of the results of \ToolName. For each single case in the clients of program slicing and bug detection, we compute the standard deviations~\cite{KitchenhamRobustStats} of precision and recall across five repeated runs, as well as the coefficients of variation~\cite{ReedCV} (i.e., the ratio of standard deviation to mean) for different computational costs. Specifically, the standard deviations of precision and recall are within 3\% across all tasks, indicating stable results under temperature 0. The remaining variability mainly stems from minor nondeterminism in greedy decoding, such as backend tie-breaking, floating-point rounding, parallelism, and scheduling on the service side. For computational costs, the coefficients of variation of input and output token costs are upper-bounded by 0.001 and 0.05, respectively. Notably, output tokens exhibit larger variation due to differences in CoT reasoning across runs, whereas input tokens are largely determined by the program and prompt templates, with only minor variation from interdependent neural relation population. Finally, the coefficient of variation of time cost is upper-bounded by 0.2, mainly due to external factors in LLM API calls (e.g., network conditions, request queueing, and server-side load).}

\smallskip
\emph{\textbf{Limitations.}}
While \ToolName\ has shown significant promise in facilitating various static analysis tasks, several limitations still need to be addressed. First, hallucination remains a persistent issue of LLMs when inferring fundamental program facts, such as alias relation upon pointers. These hallucinations, which can arise during the resolution of specific sub-tasks, may accumulate and degrade the efficacy of the overall analysis. 
Second, the high computational overhead associated with \ToolName\ could hinder its scalability for large-scale program analysis. 
When the numbers of sources and sinks in a project are large, \ToolName\ must invoke the LLM many times, which may introduce unacceptable overhead for large-scale programs.
Third, the syntax of our analysis policy language is currently restrictive. It only supports the analysis of program values induced by expressions. As a result, \ToolName\ cannot capture program properties upon other program constructs, such as functions and statements.
Also, the symbolic relations presented in Table~\ref{tab:symbolic-relation-syntax} can not support context-sensitive analysis due to the lack of support in calling context maintenance.

\smallskip
\emph{\textbf{Future Work.}}
First, fine-tuning existing LLMs using program facts derived from traditional static analyzers could improve semantic alignment, thereby reducing hallucinations. 
Second, to mitigate the overhead caused by frequent LLM prompting, smaller, task-specific code models can be employed to instantiate neural relations. For instance, if a program property, such as data dependency, is widely used across various analysis tasks, a small model could be pre-trained to efficiently populate the corresponding neural relation.
Third, we can further define additional symbolic relations, e.g., symbolic relations over program locations for context-sensitive analysis.
Typically, the users could cross-check the populated neural relations by introducing proper symbolic relations.
For example, validating the tuples in the neural relation \textsf{TaintProp} with the symbolic relation \textsf{CtrlOrd} would filter the spurious taint flows that violate control flow order.

%% file: related_work.tex
\section{Related Work}
\label{sec:related_work}

\ \ \ \ \emph{\textbf{Symbolic Static Program Analysis.}} Mainstream symbolic analyzers derive program properties from IR code generated by compilers~\cite{Xue16SVF, CadarDE08, Shi18Pinpoint, Arzt14FlowDroid} and reduce the analysis to deciding the satisfiability of constraints~\cite{CadarDE08, CalcagnoDOY09} or answering graph reachability queries~\cite{Shi18Pinpoint, Xue16SVF}. 
Typically, \textsc{FlowDroid} detects taint vulnerabilities upon Soot IR~\cite{Arzt14FlowDroid}. \textsc{Infer} constructs a sophisticated constraint system for precise memory bug detection~\cite{CalcagnoDOY09}.
While symbolic static program analyzers perform well on targeted problems, it is difficult to generalize to other tasks, such as detecting diverse bug types outside their targeted scopes. Extending them requires expertise of analysis infrastructures and even demands implementing new analysis engines~\cite{DBLP:conf/icse/JohnsonSMB13}.
For example, customizing CodeQL checkers typically requires writing over 100 lines of queries in the CodeQL query language that contains thousands of utility classes and APIs~\cite{CodeQL}.
Although lint-like checkers~\cite{DBLP:conf/ease/BennettH0C24, ClangTidy} offer compilation-free analysis, they target syntactic patterns rather than semantic ones, and thus, falling out of our scope.

\smallskip
\emph{\textbf{Static Program Analysis with Datalog.}}
Datalog-based static program analysis has gained significant traction recently~\cite{HajiyevVM06, WuZ021, YamaguchiGAR14, SmaragdakisB10}. 
Tools like \textsc{CodeQL}~\cite{AvgustinovMJS16} and \textsc{Doop}~\cite{BravenboerS09, SmaragdakisB10} formulate analysis algorithms, such as data-flow analysis and pointer analysis, with Datalog rules.
More recent studies also demonstrate the potential of domain-specific Datalog-based program analysis~\cite{DBLP:conf/pldi/BrentGLSS20, zhang2024nyx, DBLP:conf/uss/WenSCFPCD024, LiMCNWS21}.
\ToolName\ employs Datalog as a declarative analysis policy language to decompose analysis tasks into manageable sub-problems. Unlike traditional approaches, \ToolName\ allows users to customize analyses by defining neural relations through examples and natural language descriptions, rather than crafting additional rules, significantly lowering the barrier to customization.

\smallskip
\emph{\textbf{Machine Learning-based Static Program Analysis.}}
\label{subsec:mlsa}
Recent advances in machine learning have opened new opportunities for static program analysis~\cite{DBLP:conf/icse/SteenhoekGL24, DBLP:conf/ijcnn/HanifM22, DBLP:conf/iclr/DinellaDLNSW20}.
For example, \textsc{DeepDFA} trains embedding and classification models for bug detection~\cite{DBLP:conf/icse/SteenhoekGL24}.
Other studies explore foundational analysis tasks, such as equivalence checking~\cite{DBLP:journals/pacmpl/ChenYLZWWSW24} and dependency analysis~\cite{DBLP:conf/icse/YadavallyNWW23}.
Although these approaches avoid compilation, their reliance on high-quality training datasets limits their customizability.
Leveraging the LLMs' ability in code semantic understanding, recent studies introduce various neuro-symbolic designs for static analysis~\cite{DBLP:journals/pacmpl/LiHZQ24, DBLP:conf/sosp/YangZXLZ25, DBLP:journals/corr/abs-2504-16057}.
Typcially, KNighter~\cite{DBLP:conf/sosp/YangZXLZ25} and MoCQ~\cite{DBLP:journals/corr/abs-2504-16057} enable customized bug detection by synthesizing symbolic checkers using LLMs.
IRIS facilitates taint-style vulnerability detection by identifying sources and sinks with LLMs.
Although they can customize bug detection,
they primarily rely on symbolic analysis for code semantic reasoning, which restricts bug types they can handle.
For instance, they cannot detect resource exhaustion or unsalted one-way hash bugs targeted in our evaluation.
In contrast, \ToolName\ leverages the ability of LLMs to jointly understand natural languages and interpret code semantics~\cite{DBLP:journals/corr/abs-2401-00812}, enabling prompting-based customization that supports diverse bug types without compilation.

%% file: conclusion.tex
\section{Conclusion}
\label{sec:conclusion}

We present \ToolName, a relational neuro-symbolic framework that enables compilation-free and customizable static program analysis. 
Based on an analysis policy,
\ToolName\ decomposes an analysis task and employs parsing-based analysis alongside LLM prompting to reduce hallucinations.
\ToolName\ achieves a performance comparable to, and even exceeding, state-of-the-art approaches in different analysis tasks,
uncovering 13 previously unknown real-world bugs that have been confirmed and fixed.
We believe \ToolName\ offers valuable insights into the intersection of LLMs and symbolic analysis, paving the way for reshaping static analysis for better usability.